\documentclass{article}
\usepackage[utf8]{inputenc}
\usepackage[a4paper,width=170mm,top=15mm,bottom=25mm]{geometry}
\usepackage{amsmath} 
\usepackage{graphicx}
\usepackage[graphicx]{realboxes}
\usepackage{subfig}
\usepackage{longtable}
\usepackage[utf8]{inputenc}
\usepackage{caption}
\usepackage{float}
\usepackage{pdflscape}
\usepackage{xcolor}
\usepackage{stackengine}
\usepackage{gensymb}
\usepackage{fancyhdr}
\usepackage{amsmath}
\usepackage{amssymb}
\usepackage{authblk}
\usepackage{float}
\usepackage{pdfpages}
\usepackage{titlesec}
\usepackage{gensymb}

\begin{document}

\title{\textbf{CKM matrix parameters from the exceptional Jordan algebra}}

\author[1]{Aditya Ankur Patel}
\author[2]{Tejinder P. Singh}
\affil[1]{\it Indian Institute of Science Education and Research  Mohali\linebreak
\it Sector 81, S.A.S Nagar, Manauli PO, Mohali 140306, India\linebreak
\tt Email: ms18032@iisermohali.ac.in}
\affil[2]{\it Inter-University Centre for Astronomy and Astrophysics,\linebreak \it Post Bag 4, Ganeshkhind,  Pune 411007, India\linebreak
and\linebreak
\it Tata Institute of Fundamental Research,\linebreak
\it Homi Bhabha Road, Mumbai 400005, India\linebreak
\tt Email:  tejinder.singh@iucaa.in, tpsingh@tifr.res.in}
 
\date{\today}

\maketitle

\begin{abstract}
\noindent We report a theoretical derivation of the Cabibbo-Kobayashi-Maskawa (CKM) matrix parameters and the accompanying mixing angles. These results are arrived at from the exceptional Jordan algebra applied to quark states, and from expressing flavor eigenstates (i.e. left-chiral states) as superposition of mass eigenstates (i.e. the right-chiral states) weighted by square-root of mass. Flavor mixing for quarks is mediated by the square-root mass eigenstates, and the mass ratios used have been derived in earlier work from a left-right symmetric extension of the standard model. This permits a construction of the CKM matrix from first principles.
There exist only four normed division algebras, they can be listed as follows - the real numbers $\mathbb{R}$, the complex numbers $\mathbb{C}$, the quaternions $\mathbb{H}$ and the octonions $\mathbb{O}$. The first three algebras are fairly well known; however, octonions as algebra are less studied. Recent research has pointed towards the importance of octonions in the study of high energy physics.
Clifford algebras and the standard model are being studied closely. The main advantage of this approach is that the spinor representations of the fundamental fermions can be constructed easily here as the left ideals of the algebra. Also, the action of various Spin Groups on these representations too can be studied easily.
In this work, we build on some recent advances in the field and try to determine the CKM angles from an algebraic framework. We obtain the mixing angle values as $\theta_{12}=11.093^o, \theta_{13}=0.172^o, \theta_{23}=4.054^o$. In comparison, the corresponding experimentally measured values for these angles are $13.04^o \pm 0.05^o, 0.201^o \pm 0.011^o, 2.38^o \pm 0.06^o $. The agreement of theory with experiment is likely to improve when running of quark masses is taken into account. 
\end{abstract}
\tableofcontents
\section{Introduction}
There has been occasional interest in the last few decades as to the significance of octonions for understanding the standard model of particle physics \cite{9a889e59f85647a29d63164144d7583b}. Research on this topic has picked up significant pace in the last seven years or so, since the publication of Furey's Ph. D. thesis \cite{furey2016standard}, and also the discovery by Todorov and Dubois-Violette \cite{Todorov} that the exceptional groups $G_2, F_4, E_6$ contain symmetries of the standard model as maximal sub-groups. This has given rise to the hope that octonions could play a significant rule in the unification of electroweak and strong interaction, and in turn their unification with gravitation. Octonionic chains can be used to generate a Clifford algebra, and spinors made as minimal left ideals of Clifford algebras possess symmetries observed in the standard model \cite{fureycharge, furey2016standard}.

We have proposed a left-right symmetric extension of the standard model, based on complex split bioctonions, which incorporates gravitation \cite{vaibhav2021leftright}. This is consistent with unification based on an $E_8 \times E_8$ symmetry, and the breaking of this symmetry reveals the standard model \cite{kaushik2022e8, Sherry}. Chiral fermions arise after symmetry breaking; left-handed fermions are eigenstates of electric charge and right-handed fermions are eigenstates of the newly introduced $U(1)$ quantum number - square root of mass. By expressing charge eigenstates as superpositions of square-root mass eigenstates one is able to theoretically derive the observed mass ratios of quarks and charged leptons \cite{Singhfsc, bhatt2022majorana, Singhwhy, Singhexceptional, Singhgrav}.

In the present paper we extend these methods to provide a theoretical derivation of the CKM matrix parameters for quark mixing, and the accompanying mixing angles. Also, we show that the complex Clifford algebra $Cl(9)$ is the algebra of unification. Further, we conclude from our investigations that our universe possesses a second 4D spacetime with its own distinct light-cone structure. Distances in this space-time are invariably microscopic and only quantum systems can access this second space-time.

This paper is organised as follows. Sections 2, 3, 4 review a few basics of group representations, Clifford algebras, and the octonions. Sections 5 and 6 briefly recall earlier work on particle representations made from the octonions, and our own work on derivation of mass ratios from the exceptional Jordan algebra. Section 7 is the heart of the paper; the space of minimal ideals is constructed and the role of $SU(2)_L$ and $SU(2)_R$ symmetry elucidated. The triality property of the spinor and vector reps of $SO(8)$ is used to motivate the methodology for theoretical derivation of the CKM matrix parameters. The calculation of these matrix parameters and mixing angles is then carried out in Section 8. Conclusions are in Section 9.

The CKM matrix plays a central role in the understanding of weak interactions of quarks and provides a quantitative measure of the flavor change brought about by these interactions. It plays a key role in the understanding of CP violation, and a possible violation of the unitarity condition might be an indication of physics beyond the standard model. What is important is to note that to date our knowledge of the CKM matrix parameters comes exclusively from experiment. The CKM angles are free parameters of the standard model and there is no generally accepted theory which explains why these angles should have the values measured in experiments. To the best of our knowledge, the present paper is the first to provide a first principles derivation of the CKM angles, starting from a theory of unification of the standard model with gravitation. Based on spontaneous breaking of the unified $E_8 \times E_8$ symmetry, a new $U(1)$ symmetry arises, which we named $U(1)_{grav}$. Its associated charge is square root of mass $\pm\sqrt m$ which can have either sign (analogous to electric charge): positive sign for matter and negative sign for anti-matter. Left-handed fermion states are eigenstates of electric charge and right-handed fermion states are eigenstates of square-root mass. These characteristics enable us to construct the CKM matrix, and the fact that mass eigenstates are labelled by square root mass and not by mass plays a very important role in correctly determining values of the CKM angles. An earlier paper on CKM angles which foresaw the significance of square root mass is the one by Nishida \cite{Nishida} and is titled `Phenomenological formula for CKM matrix and its physical interpretation'. An even earlier interesting work is by Fritzsch \cite{F1, F2} who also aimed to derive the mixing angles in terms of quark mass ratios. While these important works bear some interesting similarity to ours, they take quark masses and their ratios as inputs from experiment. On the other hand, we have first derived mass ratios from an underlying theory of unification, and in the present work these mass ratios are used to derive the weak mixing angles. Thus the octonionic theory of unification provides strong evidence that the fundamental constants of the standard model are derivable from a coherent framework and are not free parameters of nature.

\section{A Few Basics}
To engage in the study of the Clifford algebras, mass ratios and their application to the standard model itself, we first need a basic introduction to some mathematical concepts. A basic review has been done in the following  sections about some of the required concepts.
\subsection{Algebra}
 An algebra (A, + , . , F) over a field $F$ is defined to be a vector space over the field, equipped with a bi-linear operation that follows the following properties:-
 \begin{equation}
     m : A \times A \longrightarrow A 
 \end{equation}
 \begin{equation}
     (a , b) \longrightarrow a.b \:\:\:\:\: a, b, a.b \in A
 \end{equation}
\begin{itemize}
    \item ($\alpha$a).($\beta$b) = $\alpha\beta$(a.b) \:\:\:\:\:\ $\alpha$, $\beta$ $\in$ F ; a, b $\in$ A
    \item (a + b).c = (a.c) + (b.c) \:\:\:\:\: a,b,c $\in$ A.
    \item ((a.b).c) = (a.(b.c))
\end{itemize}
An ideal $I$ is defined as a subspace of $A$ which survives multiplication by any element of $A$. A left ideal is defined as:-
\begin{equation}
a \in I , \forall b \in A \implies (b.a) \in I
\end{equation}
\subsection{Group Representations}
We recall a few essential basics about group theory.
\begin{itemize}
    \item If there is a  homomorphism from a  group $G$ to a group of operators $U(G)$ on a vector space $V$, then $U(G)$ forms a representation of group $G$ on $V$.
    \item The dimension of the representation is the same as the dimension of the vector space:
\end{itemize}
\begin{equation}
   g \in G \overset{U}{\longrightarrow} U(g)  
\end{equation}
\begin{equation}
    U(g)e_{i} = D(g)^{j}_{i}e_{j} \:\:\:\:\:\:\:\:\:\:\:\:\:\:\: i,j = 1,2 --dim(V)
\end{equation}
Here, the $D$ are the matrix representation of $G$ on the vector space $V$.
As a representation is a homomorphism it must preserve the group operation, so we have:-
\begin{equation}
    U(g_{1})U(g_{2}) = U(g_{1}.g_{2})
\end{equation}
\begin{equation}
    D(g_{1})D(g_{2}) = D(g_{1}.g_{2})
\end{equation}
If for a representation $U(G)$ of $G$ on $V$, there exists a subspace $V_{1}$ in $V$ such that :-
\begin{equation}
    U(g)|x_{1}\rangle \in V_{1} \:\:\:\:\ \forall x_{1} \in V_{1} 
\end{equation}
then such a subspace is called an invariant subspace of $V$ with respect to the group representation $U(G)$. The trivial invariant subspaces of $V$  are $V$ itself, and the space of null vectors.
A subspace which does not have any non-trivial invariant subspace is called minimal or proper.
The representation $U(G)$ on $V$ is called \textbf{irreducible} if there is no non-trivial invariant subspace in $V$; otherwise the representation is \textbf{reducible} \cite{tung1985group}.
\subsection{The Standard Model}
\begin{table}[ht!]
\centering
    \begin{tabular}{||c c c ||} 
 \hline
 \textbf{Force} & \textbf{Gauge Boson} & \textbf{Symbol}  \\ [0.5ex] 
 \hline\hline
 Electromagnetism & Photon & $\gamma$  \\ 
 \hline
 Weak Force & W and Z bosons & $W^{+} , W^{-}, Z$  \\
 \hline
 Strong Force & Gluons & g \\
 [1ex] 
 \hline
\end{tabular}
\caption{Forces and force carriers.}
\end{table}
The gauge group of the standard model is given below:
 \begin{equation}
    G_{SM} = SU(3)_{c}\times SU(2)_{L} \times U(1)_{Y}
\end{equation}
Also, the forces and their respective carriers are presented in  Table $1$.
\begin{itemize}
    \item A representation of the gauge group $G$ acts on a finite-dimensional Hilbert space $V$.
    \item Particles then live in the irreducible invariant subspace of $V$ as their \textbf{basis vectors}.
\end{itemize}
\section{Clifford Algebras}
A Clifford algebra $Cl(p,q)$ over $\mathbb{R}$ is defined to be an associative algebra, 
generated by $n$ elements $e_{i}$.
These $n$ generators exhibit the properties:
\begin{equation}
\bigl\{e_{i},e_{j}\bigl\} = e_{i}e_{j} + e_{j}e_{i} = 2\eta_{ij}
\end{equation}
\begin{equation}
    e_{i}^{2} = 1 \:\:\:\: e_{j}^{2} = -1 
\end{equation}
Here $i$ runs from $1$ to $p$, $j$ runs from $1$ to $q$.
The multiplication, also called the Clifford product, can be realised in terms of dot product and wedge product of vectors. An example is:-
\begin{equation}
    xy = \textbf{x}.\textbf{y} + \textbf{x}\wedge\textbf{y}
\end{equation}
The signature becomes irrelevant when we form the algebra over $\mathbb{C}$ as the field.
For a vector $v$ (a linear combination of generators) we have :-
\begin{equation}
    v^{2} = -\lvert \lvert v \rvert\rvert \implies   v^{-1} = \frac{-v}{\lvert \lvert v \rvert\rvert} 
\end{equation}
\subsection{Pin and Spin Groups}
There is a natural automorphism in the Clifford algebra, for all vectors in the Clifford algebra, given by:-
\begin{eqnarray}
v\longrightarrow \Tilde{v} = -v 
\end{eqnarray}
Let us denote this automorphism as $\alpha$.
It partitions the algebra into two parts, firstly we have the part that is the product of even number of vectors, given as:
\begin{equation}
    Cl^{even}(n) = \biggl\{\alpha(x) = x; \forall x \in Cl(n) \biggl\} 
\end{equation}
The other part contains odd number of vectors as product
\begin{equation}
    Cl^{odd}(n) = \biggl\{\alpha(x) = -x; \forall x \in Cl(n) \biggl\} 
\end{equation}
For a non-null vector $u$, we can define an inverse given by:-
\begin{eqnarray}
\forall u \in V \subset Cl(V)\\
\exists u^{-1} \in Cl^{*}(V) : u^{-1} = -\frac{u}{\lVert \Vec{u} \rVert}
\end{eqnarray}
Here  $Cl^{*}(V)$ is the group of elements that have inverses. The definition of inverse of vector can be extended to the inverse of the product of the vectors. Thus, we can define two groups as done below \cite{gallier2014clifford, harvey1990spinors}:-
\begin{equation}
    \textbf{Pin} = \left\{ a \in Cl^{*}(V) : a = u_{1}u_{2}---u_{r} , u_{j} \in  V , |u_{j}|= 1\right\}
\end{equation}
\begin{equation}
    \textbf{Spin} = Pin \cap Cl^{even}(n) =\biggl\{  a \in Cl(n) :a = u_{1}-u_{2r} ; u_{j}\in V , \lvert u_{j} \rvert = 1 \biggl\}
\end{equation}

The action of both these groups on $V$ can be defined by the \textbf{Twisted Adjoint Action}:-
\begin{equation}
    \tilde {Ad}_{a}x = \alpha(a)xa^{-1} \in V \:\:\:\:\: \forall x \in V.
\end{equation}
\begin{equation}
    (\alpha(a)va^{-1})^{2} = v^{2} \:\:\:\:\:\: \forall v \in V
\end{equation}
As both these group preserve the magnitude of the vectors, they are \textbf{orthogonal} and \textbf{special orthogonal} transformations.
\begin{equation}
\textbf{Pin} \longrightarrow O(n)
\end{equation}
\begin{equation}
\textbf{Spin} \longrightarrow SO(n)
\end{equation}
\subsection{Representations of Clifford Algebras}
The real and complex Clifford algebras have matrix representations. Here however, we will focus on representations of complex Clifford algebras.
The representations of the even subalgebra can be similarly obtained by the  identity \cite{baez2002octonions, 10.1093/ptep/ptaa104}:-
\begin{equation}
    Cl^{even}(n) \cong Cl(n-1) \:\:\:\ ; n \geq 1
\end{equation}
The matrix representations are given below, here $M_{p}(\mathbb{C})$ represents a $p \times p$ matrix with complex entries.
\begin{equation}
    Cl(n) \cong M_{p}(C) \:\:\:\:\:\:\:\:\   p = 2^{\frac{n}{2}} \:\:\:\ ; n = even  
\end{equation}
\begin{equation}
    Cl(n) \cong M_{p}(C) \oplus M_{p}(C) \:\:\:\:\:\:\:\:\   p = 2^{\frac{n-1}{2}} \:\:\:\ ; n = odd 
\end{equation}
Again notice that for the odd case, the total representation gets reduced to two irreducible representations. Especially look at the case of $n = 3,7 \ {\rm mod}\ 8$. The irreducible subspace on which matrices act is represented by $P$. These $M_{n}(F)$ act on n-dimensional irreducible space. The choice of volume element can split the algebra into two parts \cite{todorov2011clifford, 10.1093/ptep/ptaa104}, total space also gets partitioned into two irreducible subspaces. For dimensions 3 and 7, there are two choices of the irreducible spaces, \textbf{positive pinor space} ($P_{+}$) and \textbf{negative pinor space} ($P_{-}$).
\begin{equation}
    Cl(n) \cong Cl^{+}(n) \oplus Cl^{-}(n) \cong End_{C}(P_{+}) \oplus End_{C}(P_{-}) 
\end{equation}
\begin{equation}
    P = P_{+} \oplus P_{-}
\end{equation}
Now look at the case for the complexified Dirac algebra $\mathbb{C}\otimes Cl(1,3)$, it is equivalent to complex Clifford algebra $Cl(4)$. We need to study the usual spinors, so we look at the matrix representations of $Cl^{even}(4)$. We know that $Cl^{even}(4) \cong Cl(3)$. For those cases, where the even subalgebra gets partitioned into two, we similarly get \textbf{positive spinor space ($S_{+}$)} and \textbf{negative spinor space ($S_{-}$)}.
\begin{equation}
    Cl^{even}(4) \cong Cl(3) \cong M_{2}(C)\oplus M_{2}(C)
\end{equation}
\begin{equation}
    S = S_{+} \oplus S_{-} = S_{L} \oplus S_{R}
\end{equation}
These are the matrix representations of the spin groups that act on the spinor space. The total spinor space is the vector sum of the positive and negative spinor spaces. Both spaces are 2 dimensional and indeed these spaces are interpreted as the \textbf{left handed Weyl spinor} and \textbf{right handed Weyl spinor}. 
Keeping this information in mind, we will construct two irreducible subspaces in higher dimensions, brief outline is discussed below. For the $Cl(8)$ algebra we look at its even subalgebra.
\begin{equation}
    Cl^{even}(8) = Cl(7) = M_{8}(\mathbb{C})\oplus M_{8}(\mathbb{C})
\end{equation}
As $n = 7$, the representation space can be decomposed into two irreducible subspaces. This fact can be used later to include spin and other things in the analysis.
\section{Octonions}
A generic complex octonion can be represented as :
\begin{equation}
    \mathbb{C}\otimes\mathbb{O} = \sum_{n = 0}^{7}A_{n}e_{n}
\end{equation}
Here $A_{n}$ are complex coefficients and $e_{n}$ are octonionic units, with properties $e_{0}^{2} = 1$ and $e_{i}^2 = -1$. So $e_{0} = 1$ and rest are the imaginary octonionic units.
In general octonionic multiplication is non-associative; an example is given:
\begin{equation}
e_{3}(e_{4}(e_{6} + ie_{2})) = -1 + ie_{7}
\end{equation}
\begin{equation}
    (e_{3}e_{4})(e_{6} + ie_{2}) = -1-ie_{7}
\end{equation}
To tackle this problem of the octonions, we need to define an order of multiplication on a product of octonions. It leads to chain of octonions, made from maps.
\begin{equation}
    e_{1}(e_{2}(e_{3}(e_{4})))) \longrightarrow \overleftarrow{e_{1}e_{2}e_{3}e_{4}}
\end{equation}
\begin{equation}
    \overleftarrow{--e_{i}e_{j}--}f = -(\overleftarrow{--e_{j}e_{i}--}f)
\end{equation}
We will work with octonionic chains only. Octonionic multiplication is represented by the Fano Plane given below. An example multiplication is given by:-
\begin{equation}
    e_{7}e_{1} = e_{3} \:\:\:\: {\rm and} \:\:\:\:e_{1}e_{7} = -e_{3}
\end{equation}
\begin{equation}
    e_{i}e_{j} + e_{j}e_{i} = 0
\end{equation}
The octonionic chains form a representation of the Clifford algebras and hence we are interested in their study. They form a representation of $Cl(6)$ \cite{furey2016standard}. 
\begin{figure}[ht!]
    \centering
    \includegraphics[width=7cm]{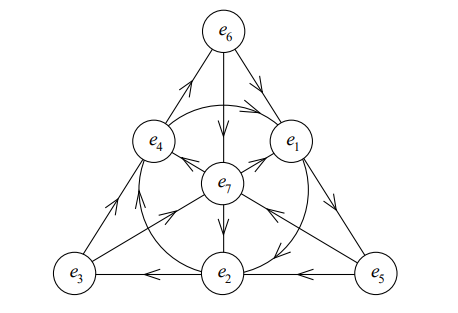}
    \caption{The Fano Plane \cite{baez2002octonions}.}
\end{figure}
The  generators of the Clifford algebra can be constructed from the octonionic imaginary units as shown in Furey's work \cite{furey2016standard}. The Fano plane in Figure $1$, lists down the method to multiply octonionic units.

\begin{equation}
    Cl(6) \cong \mathbb{C}\otimes \overleftarrow{\mathbb{O}}
\end{equation}
 The 64 dim $Cl(6)$ algebra is fully generated by the set : $\big\{$ $\overleftarrow{ie_{1}}, \overleftarrow{ie_{2}}, \overleftarrow{ie_{3}}, \overleftarrow{ie_{4}},\overleftarrow{ie_{5}},\overleftarrow{ie_{6}}$ $\bigl\}$. These are the generators of the Clifford algebra and  act as the underlying vector space structure.
\begin{equation}
    \overleftarrow{e_{1}e_{2}e_{3}e_{4}e_{5}e_{6}} f = \overleftarrow{e_{7}}f
\end{equation}
\section{Minimal Left Ideals}
The generators of $Cl(6)$ can be used to make elements of maximally totally isotropic space (MTIS). An element of maximally totally isotropic space has a quadratic norm equal to zero \cite{Ablamowicz1995}. This space for Maximally Isotropic Subspace follows the algebraic structure given below:-
\begin{equation}
    \bigl\{ q_{i},q_{j} \bigl\}f = q_{i}(q_{j}f) + q_{j}(q_{i}f) = 0 
\end{equation}
\begin{equation}
    \bigl\{ q_{i}^{\dagger},q_{j}^{\dagger} \bigl\}f = q_{i}^{\dagger}(q_{j}^{\dagger}f) + q_{j}^{\dagger}(q_{i}^{\dagger}f) = 0 
\end{equation}
\begin{equation}
     \bigl\{ q_{i},q_{j}^{\dagger} \bigl\}f = \delta_{ij}f
\end{equation}
The $a^{\dagger}$ represents  Hermitian conjugation. It is basically the complex conjugation $a^{*}$ and octonionic conjugation $\Tilde{a}$ done simultaneously. The elements of the MTIS  can be constructed from the generators of $Cl(6)$. One choice is given below \cite{furey2016standard, vaibhav2021leftright}. The six generators give rise to six elements with a quadratic norm equal to zero. There can be other equivalent choices also \cite{Ablamowicz1995}
\begin{equation}
    q_{1} = \frac{1}{2}(-e_{5} + ie_{4}) \:\:\:\:\:\:\:\  q_{1}^{\dagger} = \frac{1}{2}(e_{5} + ie_{4}) 
\end{equation}
\begin{equation}
    q_{2} = \frac{1}{2}(-e_{3} + ie_{1})\:\:\:\:\:\:\:\   q_{2}^{\dagger} = \frac{1}{2}(e_{3} + ie_{1})
\end{equation}
\begin{equation}
    q_{3} = \frac{1}{2}(-e_{6} + ie_{2}) \:\:\:\:\:\:\:\  q_{3}^{\dagger} = \frac{1}{2}(e_{6} + ie_{2}) 
\end{equation}
We construct quantities out of these isotropic vectors, the \textbf{nilpotent} given as \cite{furey2016standard}:-
\begin{equation}
   q  = q_{1}q_{2}q_{3} \:\:\:\:\:\:\:\:\:\ q^{\dagger} = q_{3}^{\dagger}q_{2}^{\dagger}q_{1}^{\dagger}
\end{equation}
\begin{equation}
    q^{2} =  0  \:\:\:\:\:\:\:\:\:\ (q^{\dagger})^{2} = 0
\end{equation}
We also have the \textbf{idempotent} given as:-
\begin{equation}
   p = qq^{\dagger} \:\:\:\:\:\:\:\:\:\ p^{\prime} = q^{\dagger}q
\end{equation}
\begin{equation}
    p^{2} = p \:\:\:\:\:\:\:\:\:\  (p^{\prime})^{2} = p^{\prime}
\end{equation}
We act on idempotent by the $q$ and $q^{\dagger}$ operators and get various algebraic states and the minimal left ideals. These states are later classified according to the transformations they undergo \cite{furey2016standard, Stoica_2018}.
\subsection{Symmetry Transformations}
We first look at the transformations of such kind that maximally isotropic space is closed.
Operator transforms of type :-
\begin{equation}
    e^{i\phi_{k}g_{k}} | e^{-i\phi_{k}g_{k}} \:\:\:\:\: \phi_{k} \in R
\end{equation}
\begin{equation}
    [g_{k}, \sum_{i}b_{i}q_{i}] = \sum_{j}c_{j}q_{j} \:\:\:\:\:  [g_{k}, \sum_{i}b_{i}^{\prime}q_{i}^{\dagger}] = \sum_{j}c_{j}^{\prime}q_{j}^{\dagger}
\end{equation}
We can make Hermitian operators by the following procedures:-
\begin{equation}
    q = c_{1}q_{1} + c_{2}q_{2} + c_{3}q_{3} \:\:\: and \:\:\: q^{\prime} =  c_{1}^{\prime}q_{1} + c_{2}^{\prime}q_{2} + c_{3}^{\prime}q_{3}
\end{equation}
Charge operator has a $U(1)$ symmetry: Q = $\frac{1}{3}\sum_{i} q^{\dagger}_{i}q_{i}$, and 
SU(3) generators : 
\begin{equation}
\Lambda_{1} = -q_{2}^{\dagger}q_{1} - q_{1}^{\dagger}q_{2} \:\:\:\:\:  \Lambda_{2} = iq_{2}^{\dagger}q_{1} - iq_{1}^{\dagger}q_{2}
\end{equation}
\begin{equation}
\Lambda_{3} = q_{2}^{\dagger}q_{2} - q_{1}^{\dagger}q_{1} \:\:\:\:\: \Lambda_{4} = -q_{1}^{\dagger}q_{3} - q_{3}^{\dagger}q_{1}
\end{equation}
\begin{equation}
\Lambda_{5} = -iq_{1}^{\dagger}q_{3} + iq_{3}^{\dagger}q_{1} \:\:\:\:\: \Lambda_{6} = -q_{3}^{\dagger}q_{2} -q_{2}^{\dagger}q_{3}
\end{equation}
\begin{equation}
\Lambda_{7} = iq_{3}^{\dagger}q_{2} - iq_{2}^{\dagger}q_{3} \:\:\:\:\: \Lambda_{8} = -\frac{1}{\sqrt{3}}(q_{1}^{\dagger}q_{1} + q_{2}^{\dagger}q_{2} - 2q_{3}^{\dagger}q_{3})
\end{equation}
A general Hermitian operator can be written as :
\begin{equation}
    \sum_{H} H = r_{0}Q + r_{i}\sum_{i = 1}^{8}\Lambda_{i}
\end{equation}
We see that the idempotent remains unaffected by these operations:
\begin{equation}
e^{i\sum H}qq^{\dagger}e^{-i\sum H} = (1 + i\sum H+--)qq^{\dagger}(1 - i\sum H --) = qq^{\dagger} = p
\end{equation}
Hence it is identified as a neutrino. 
The down isospin family can be obtained via complex conjugation of all the particles. Operators for that family also get complex conjugated and then are used to identify the particles.
\subsection{Particle Representations}
We have the  symmetry groups $SU(3)$ and $U(1)$ of the standard model; we now  look at the action of these groups on the elements of the minimal left Ideals and see how they transform. Depending upon their transformations and eigenvalues we label them accordingly \cite{furey2016standard}, as shown in Table 2.
We look at their charges obtained by the action of the $Q$ operator and  also observe the action of $SU(3)$ generators to classify them.
\begin{table}[ht!]
\centering
    \begin{tabular}{||c c c c||} 
     \hline
     $Q$ & $\Lambda$ & $P^{u}$ &  Particle\\ [0.5ex] 
     \hline\hline
     0 & 1 &  $p$& $\nu$  \\ 
     \hline
     $\frac{1}{3}$ & $\Bar{3}$ & $q_{i}^{\dagger}p$ & $\Bar{d_{i}}$ \\ 
     \hline
     $\frac{2}{3}$ & 3 & $q_{i}^{\dagger}q_{j}^{\dagger}p$ & $u_{i}$ \\ 
     \hline
     1 & 1 & $q_{i}^{\dagger}q_{j}^{\dagger}q_{k}^{\dagger}p$ & $e^{+}$ \\ 
     \hline
    \end{tabular}
    \begin{tabular}{||c c c c||} 
     \hline
     $-Q^{*}$ & $-\Lambda^{*}$ & $P^{d}$ &  Particle\\ [0.5ex] 
     \hline\hline
     0 & 1 &  $p^{\prime}$& $\Bar{\nu}$ \\ 
     \hline
     $-\frac{1}{3}$ & 3 & $q_{i}p^{\prime}$ & $d_{i}$ \\ 
     \hline
     $-\frac{2}{3}$ & $\Bar{3}$ & $q_{i}q_{j}p^{\prime}$ & $\Bar{u_{i}}$ \\ 
     \hline
     $-1$ & 1 & $q_{i}q_{j}q_{k}p^{\prime}$ & $e^{-}$ \\ 
     \hline
    \end{tabular}
\caption{(a)Up-Isospin particles ; (b)Down-Isospin Particles.}
\end{table}\\
The $\overline{d}_{i}$ and $u_{j}$ have indices running from 1 to 3, representing the three coloured up and anti-down quarks. The left ideal present  above gives another left ideal after the complex conjugation. This time it gives isospin down family. Observe that transition from one family to other can be done by the complex conjugation. Now the creation operator and the annihilation operator reverse their roles, we also get a new idempotent. 

Hence we have a representation of one generation of standard model particles under the unbroken symmetry $SU(3)_c\times U(1)_{em}$ \cite{furey2016standard}.
\section{Split  Bioctonions and Mass Ratios}
Split bioctonions are simply two copies of octonions in the same algebra. They can be constructed from the generators in the $Cl(7)$ algebra \cite{vaibhav2021leftright}

\begin{equation}
    Cl(7) \cong Cl(6)\oplus Cl(6) 
\end{equation}
Observe that the spinor representations of $Cl^{even}(8)$ again give us the positive and negative spinor space  
\begin{equation}
    Cl^{even}(8)\cong Cl(7) \cong M_{8}(\mathbb{C})\oplus M_{8}(\mathbb{C})
\end{equation}
\subsection{Construction}
The seven generators of Cl(7) given as : $\left\{ e_{1}, e_{2} ,e_{3}, e_{4}, e_{5},e_{6} ,e_{7} \right\}$ can be arranged in the manner  given below; keeping in mind the non-associativity of the octonions we will use the octonionic chains \cite{vaibhav2021leftright}:-
\begin{equation}
    \omega = \overleftarrow{e_{1}e_{2}e_{3}e_{4}e_{5}e_{6}e_{7}}
\end{equation}
\begin{equation}
    e_{8} = \overleftarrow{e_{1}e_{2}e_{3}e_{4}e_{5}e_{6}}
\end{equation}
\begin{equation}
    (1,e_{1},e_{2},e_{3}, e_{4}, e_{5}, e_{6}, e_{8}) \oplus \omega(1,-e_{1},-e_{2},-e_{3}, -e_{4}, -e_{5}, -e_{6}, -e_{8})
\end{equation}
\begin{equation}
    \omega^{2} = 1
\end{equation}
This $e_{8}$ acts as an octonionic unit and $\omega$ as a pseudoscalar that commutes with every octonionic unit, and hence with every element of the $Cl(7)$ algebra. It is the analog of the split complex number which squares to one but is neither one not minus one.
To generate the system with opposite parity look at an example given below:-
\begin{equation}
    \overleftarrow{e_{1}e_{2}e_{4}e_{5}e_{6}e_{7}} = -\overleftarrow{e_{1}e_{2}e_{3}e_{3}e_{4}e_{5}e_{7}e_{6}e_{6}}=-\overleftarrow{e_{3}e_{1}e_{2}e_{3}e_{4}e_{5}e_{6}e_{7}} = -e_{3}\omega = -\omega e_{3}
\end{equation}
A $Cl(6)$ algebra can be used to construct a left sided ideal. It is similar to an irreducible space; action (left multiplication) of various elements of algebra on the elements in ideal keeps the space of ideal closed, similar to the working of  irreducible space.
The two sets of octonions can now be used to construct ideals that represent  states of opposite chirality, similar to positive and negative spinor states. By the complex conjugation of the two chiral families we can also construct the antiparticle sates.
We can do so by defining the idempotents and nilpotents as earlier and do our analysis. But notice this time for the second copy of the octonions the generators have a negative sign. This helps us to introduce the chirality into the problem.
From the first copy of the octonions we get the \textbf{left handed neutrino family} and its right handed anti-particle's family.
Similarly from the second copy we can get \textbf{right handed neutrino family} and its left handed anti-particle's family \cite{vaibhav2021leftright}.
\subsection{Mass Ratios}
 We construct all three families from a single real octonionic family by a set of transformations. Both cases for Dirac and Majorana neutrino have been analysed \cite{bhatt2022majorana}. The solution of Dirac Equation in $(9,1)$ spacetime, is connected with the eigenvalue problem of the Hermitian octonionic matrices as explained in \cite{dray1999exceptional, Manogue_2010}. The eigenvalues thus calculated give us square root mass ratios of various fundamental fermions.
  \subsubsection{Hermitian Octonionic Matrices}
 The quarks have different representations for different colours. Octonions are difficult to work with, quaternions are  much easier to deal with. To make the problem simpler we take the representations of neutrino and electron and choose the colour state of quarks accordingly; such that only one quaternionic copy is used for one family of the fermions. Now this complex quaternionic representation  is mapped to real octonionic representation  by a mapping given below \cite{bhatt2022majorana}:-
 \begin{equation}
     C \otimes H \longrightarrow R\otimes O
 \end{equation}
 \begin{equation}
     (a_{0} + ia_{1}) + (a_{2} + ia_{3})e_{4} + (a_{4} + ia_{5})e_{5} + (a_{6} +ia_{7})e_{7}
 \end{equation}
 \begin{equation}
     \big\downarrow
 \end{equation}
\begin{equation}
     a_{0} + a_{1}e_{1} + a_{5}e_{2} + a_{3}e_{3} + a_{2}e_{4} + a_{4}e_{5} + a_{7}e_{6} + a_{6}e_{7}
 \end{equation}
 Once we have the real representation for one family, we do an internal rotation about some axis and  get real octonionic representation for all the three families. We can use these representations to fill the entries in $3 \times 3$ octonionic Hermitian matrices. The uniqueness of  the axis used for transformation and similar matters are already discussed earlier \cite{bhatt2022majorana}.
 It is observed that the ratios of the square root mass of the positron, the up quark and the down quark is 1 : 2 : 3. Motivated by this information we can define a new quantity as the \textbf{gravi charge}. The ratio of gravi charges will then be:-
 \begin{equation}
     e^{+} : u : d = \frac{1}{3} : \frac{2}{3} : 1
 \end{equation}
 The gravi- charges can be negative also. 
 These gravi - charges are then used on the diagonals of these octonionic Hermitian matrices. These $3 \times 3$ octonionic Hermitian matrices are referred as \textbf{exceptional Jordan matrices} and they form the exceptiomal Jordan algebra, with a specified Jordan product \cite{dray1998octonionic}
 \begin{equation}
     A o B = \frac{1}{2}(AB + BA)
 \end{equation}
 We fill the entries in the matrices accordingly with the diagonals filled with the Gravi-charge
 \begin{equation}
 X_{\nu} = 
     \begin{bmatrix}
0 & V_{\tau} & \overline{V}_{\mu}\\
\overline{V}_{\tau} & 0 & V_{\nu}\\
V_{\mu} & \overline{V}_{\nu} & 0
\end{bmatrix} \:\:\:\:\:\:\:\:\:\:\:\:\:\:\:\:
X_{e} = 
     \begin{bmatrix}
\frac{1}{3} & V_{\overline{\tau}} & \overline{V}_{\overline{\mu}}\\
 \overline{V}_{\overline{\tau}} & \frac{1}{3} & {V}_{e^{+}}\\
V_{\overline{\mu}} & \overline{V}_{e^{+}} & \frac{1}{3}
\end{bmatrix}
 \end{equation}
 \begin{equation}
     X_{u} = \begin{bmatrix}
\frac{2}{3} & V_{t} & \overline{V}_{c}\\
 \overline{V}_{t} & \frac{2}{3} & V_{u}\\
V_{c} & \overline{V}_{u} & \frac{2}{3}
\end{bmatrix} \:\:\:\:\:\:\:\:\:\:\:\:\:\:\:\:
X_{d} = \begin{bmatrix}
    1 & V_{\overline{b}} & \overline{V}_{\overline{s}}\\
    \overline{V}_{\overline{b}} & 1 & V_{\overline{d}}\\
    V_{\overline{s}} & \overline{V}_{\overline{d}} & 1
\end{bmatrix}
 \end{equation}
 These matrices satisfy the characteristic equation given as \cite{dray1998octonionic} :-
 \begin{equation}
     A^{3} - (trA)A^{2} + \sigma(A) A - (det A)I = 0
 \end{equation}
 The definition and explanation for each quantity are presented in appendix A. The exact nature of these matrices in the context of standard model is still not completely understood. However some results do suggest that the $\mathbb{OP}^{2}$ space is crucial for our understanding of the spinors, and these spaces are closely related to these Hermitian matrices \cite{boyle2020standard}.
These matrices with real octonionic entries can be further decomposed as given \cite{dray1999exceptional}:
\begin{equation}
    A = \sum_{i}^{3}  \lambda_{i}P_{\lambda_{i}}
\end{equation}
\begin{equation}
    P_{\lambda_{i}}oP_{\lambda_{j}} = 0 = \frac{1}{2}(P_{\lambda_{i}}P_{\lambda_{j}} + P_{\lambda_{j}}P_{\lambda_{i}}) 
\end{equation}
\begin{equation}
    \implies AoP_{\lambda} = \lambda P_{\lambda}
\end{equation}
It gives us an eigenmatrix equation. These eigenvalues are used to calculate the square root masses of various fundamental fermions \cite{bhatt2022majorana}, as shown in the Figure $2$ below.
\begin{figure}[ht!]
    \centering
    \includegraphics[width=10cm]{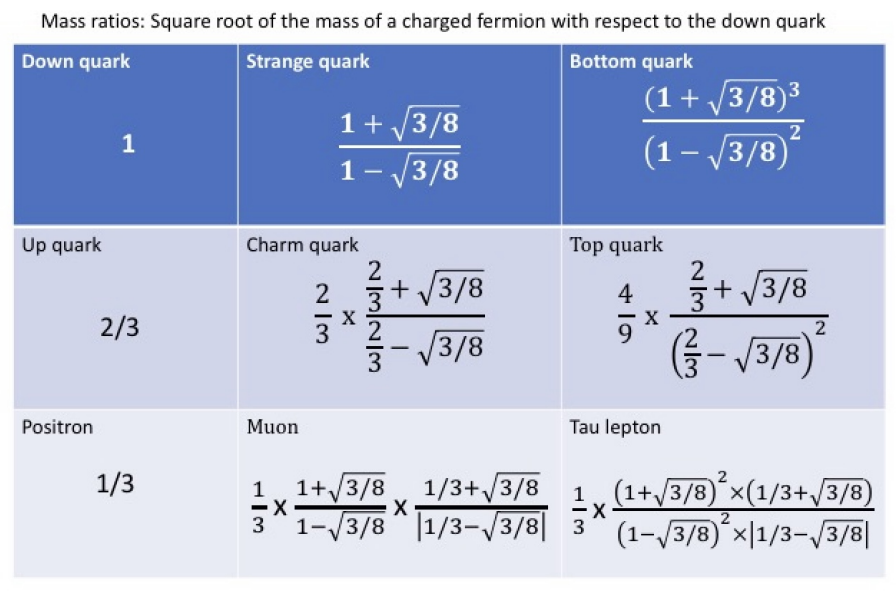}
    \caption{The square root of mass of fermions with respect to the down quark.\cite{bhatt2022majorana}}
\end{figure}
\subsubsection{Inclusion of Gravity}
The mass ratios of the up-quark, down-quark, and positron motivated us to extend the gauge group to $SU(3)_{grav}\times SU(2)_{R} \times U_{g}(1)$. This U(1) symmetry is similar to the usual $U(1)$, with \textbf{gravi-charge} as the quantity analogous to the electric charge. We can group the particles with up-isospin together as done earlier and proceed as given below:-
\begin{equation}
    e^{+} : u : \overline{d} = \frac{1}{3} : \frac{2}{3} : 1
\end{equation}
We have the following  families that are expected to observe the $SU(2)_{R}$ symmetry:-
\begin{equation}
    \begin{pmatrix}
u\\
e^{-}
\end{pmatrix} \:\:\:\:\:\:\
\begin{pmatrix}
\nu_{e}\\
d
\end{pmatrix}
\end{equation}
Notice the swapping of the down quark and electron. This structure can be extended to all  three generations.
Now we are working in the $Cl(7)$ algebra, it has two copies of the $Cl(6)$ algebra; one copy can be used to construct the octonionic representations of the gravitationally inactive  particles that transform according to the normal standard model gauge group. The second copy of the $Cl(6)$ can be used to construct a new minimal left ideal for this new extension to the gauge group, which will then have the following octonionic representation for the various gravitationally active particles. The minimal left ideal and the right handed nilpotents, and the idempotent for these spinors that are gravitationally active are then given below:-
\begin{equation}
    q_{1} = \frac{-\omega}{2}(-e_{5} + ie_{4}) \:\:\:\:\:\:\:\  q_{1}^{\dagger} = \frac{-\omega}{2}(e_{5} + ie_{4}) 
\end{equation}
\begin{equation}
    q_{2} = \frac{-\omega}{2}(-e_{3} + ie_{1})\:\:\:\:\:\:\:\   q_{2}^{\dagger} = \frac{-\omega}{2}(e_{3} + ie_{1})
\end{equation}
\begin{equation}
    q_{3} = \frac{-\omega}{2}(-e_{6} + ie_{2}) \:\:\:\:\:\:\:\  q_{3}^{\dagger} = \frac{-\omega}{2}(e_{6} + ie_{2}) 
\end{equation}
\begin{equation}
   q_{R}  = q_{1}q_{2}q_{3} \:\:\:\:\:\:\:\:\:\ q^{\dagger}_{R} = q_{3}^{\dagger}q_{2}^{\dagger}q_{1}^{\dagger}
\end{equation}
\begin{equation}
    p_{R} = q_{R}q_{R}^{\dagger}
\end{equation}\\
This helps  us to generate the following particle eigenstates :-
\begin{equation}
    \nu_{e,R} = \frac{ie_{8} + 1}{2} 
\end{equation}
\begin{equation}
    V_{{e^{+}}{1}} = \omega \frac{(-e_{5} - ie_{4})}{2} 
\end{equation}
\begin{equation}
     V_{{e^{+}}{2}} =\omega  \frac{(-e_{3} - ie_{1})}{2} 
\end{equation}
\begin{equation}
     V_{{e^{+}}{3}} =\omega  \frac{(-e_{6} - ie_{2})}{2} 
\end{equation}
\begin{equation}
    V_{{u}_{1}} = \frac{e_{4} + ie_{5}}{2} 
\end{equation}
\begin{equation}
    V_{{u}_{2}} = \frac{e_{1} + ie_3}{2} 
\end{equation}
\begin{equation}
    V_{{u}_{3}} = \frac{e_{2} + ie_6}{2} 
\end{equation}
\begin{equation}
    V_{\overline{d}} = \omega \frac{(i + e_{8})}{2}
\end{equation}
\section{Space of Minimal Left Ideals}
The complete space related to minimal left ideals is not used in the $Cl(6)$ algebra. We intend to use it fully. We already have information about the square root mass ratios. We know that $p = qq^{\dagger}$ is idempotent, and $q_{i}^{\dagger}$  are the ladder operators. By using this we can construct a left ideal and by the right multiplication on this space of left ideal we can span the whole space of the algebra \cite{Stoica_2018, gillard2019cell8}.
\subsection{Patterns in The Standard Model}
To study the standard model the first thing to do is to introduce vector spaces (or the Hilbert space) which are later made into an algebra. The underlying complex vector space $(V, h)$ establishes a natural isomorphism between vector space dual and its conjugate. $h$ here is the inner product on the vector space. We, therefore  have the following relations \cite{Stoica_2018}:- 
\begin{eqnarray}
V^{-1} \cong V^{\dagger} \cong \overline{V} 
\end{eqnarray}
\begin{table}[ht!]
\centering
 \begin{tabular}{||c c c c||} 
 \hline
 Force/Charge & Internal space & Dimension & Symmetry \\ [0.5ex] 
 \hline\hline
 Electromagnetism & $\chi_{em}$ & 1 & U(1) \\ 
 \hline
 Strong & $\chi_{c}$ & 3 & SU(3) \\
 \hline
 Weak Hypercharge & $\chi_{Y}$ & 1 & U(1) \\
 \hline
 Weak- Electromagnetism & $\chi_{ew}$ & 2 & U(2) \\ [1ex] 
 \hline
\end{tabular}   
\caption{Internal Space for various symmetries.}
\end{table}\\
Table $3$ represents the vector space required to explain the appropriate symmetries \cite{Stoica_2018}.
The space $\chi_{em}$ represents one vector that corresponds to a charge of $\frac{e}{3}$ and the space $\chi_{c}$ represents a 3-dimensional complex vector space that have three basis vectors given as  $\{ r , g, b\}$. For the electromagnetic space the charges add up for the tensor product of such spaces, they appear as \textbf{numbers in the exponential associated with the U(1) symmetry}. By the above relations we then have information about the dual space or the conjugate space. We have the space $\overline{\chi}_{em}$ which has the charge equal to $\frac{-e}{3}$; and the dual colour space which now has the vectors as  $\{ \overline{r} , \overline{g}, \overline{b}\}$.
We can use our knowledge of how particles transform under various symmetry transformations and define the internal elctro-colour space for various particles as done in Table $4$. This will later help us to develop isomorphisms between exterior algebra related to the internal space and the elements of the $Cl(6)$ algebra. 
\begin{table}[ht!]
\centering
    \begin{tabular}{||c  c||} 
 \hline
 Particle & Internal Space \\ [0.5ex] 
 \hline\hline
 $e^{-}$ & $\chi_{em}^{3}$ \\ 
 \hline
 $\overline{u}$ & $\chi_{em}^{2}\overline{\chi}_{c}$ \\
 \hline
 d & $\chi_{em}\chi_{c}$ \\
 \hline
 $\nu$ & C \\
 \hline
 $\overline{\nu}$ & C \\
 \hline
 $\overline{d}$ & $\chi_{em}^{-1}\overline{\chi}_{c}$ \\
 \hline
 u & $\chi_{em}^{-2}\Bar{\chi_{c}}$ \\
 \hline 
 $e^{+}$ & $\chi_{em}^{-3}$ \\ [1ex] 
 \hline
\end{tabular}
    
\caption{Internal Space of Particles.}
\end{table}
For the colour space of fermions, we can use the exterior powers of the $\chi_{c}$ to represent different fermions. The colour space $\chi_{c}$ and its dual (or conjugate) $\overline{\chi}_{c}$ have the basis as given below:-
\begin{equation}
    \chi_{c} = \{ r,g,b \} \:\:\:\:\:\:\:\:\:\:\:\:\:\:\ \overline{\chi}_{c} = \{ \overline{r} , \overline{g} , \overline{b} \}
\end{equation}
For the exterior algebra of the vectors of the colour space and its dual, we have the following relation:-
\begin{equation}
    \Lambda^{-k}\chi_{c} = \Lambda^{k}\overline{\chi}_{c}
\end{equation}
With this knowledge we have the following isomorphisms:-
\begin{equation}
    \Lambda^{0}\chi_{c} \cong \mathbb{C}
\end{equation}
\begin{equation}
    \Lambda^{1}\chi_{c} \cong \chi_{c}
\end{equation}
\begin{equation}
    \Lambda^{2}\chi_{c} \cong \overline{\chi}_{c} : \:\:\:\:\:\:\:\:\ \{ r \wedge g \rightarrow \overline{b} \:\:\:\ , \:\:\:\ g \wedge b \rightarrow \overline{r} \:\:\:\ , \:\:\:\ r \wedge b \rightarrow \overline{g} \}
\end{equation}
\begin{equation}
    \Lambda^{3}\chi_{c} \cong \mathbb{C} :  \:\:\:\:\:\:\:\:\    r \wedge g \wedge b
\end{equation}
The representations of particles in exterior algebra are given in Table $5$.
For the simplification of the notation define :-
\begin{equation}
    \chi= \overline{\chi}_{em}\otimes \overline{\chi}_{c} 
\end{equation}
Note that the Hilbert space is equipped with $h = h_{em}\otimes h_{c}$, and the space is 3 dimensional.\\
\begin{table}[ht!]
\centering
    \begin{tabular}{||c  c||} 
 \hline
 Particles & Vectors in Exterior Space \\ [0.5ex] 
 \hline\hline
 $e^{-}$ & $\Lambda^{3}\overline{\chi}$ \\ 
 \hline
 $\overline{u}$ & $\Lambda^{2}\overline{\chi}$ \\
 \hline
 d & $\Lambda^{1}\overline{\chi}$ \\
 \hline
 $\nu$ & $\Lambda^{0}\overline{\chi}$ \\
 \hline
 $\overline{\nu}$ & $\Lambda^{0}\chi$ \\
 \hline
 $\overline{d}$ & $\Lambda^{1}\chi$ \\
 \hline
 u & $\Lambda^{2}\chi$  \\
 \hline 
 $e^{+}$ & $\Lambda^{3}\chi$  \\ [1ex] 
 \hline
\end{tabular}
\caption{Particles as the representations of the Exterior Algebra.\cite{Stoica_2018}}
\end{table}\\
We choose a basis of the isotropic vectors for the newly defined space $\chi$ as $ \{ q_{1} ,q_{2} , q_{3} \}$, and its dual basis for the space $\overline{\chi}$ as $ \{ q_{1}^{\dagger} ,q_{2}^{\dagger} , q_{3}^{\dagger} \}$. So the total Hilbert space can be seen as $\chi^{\dagger} \oplus \chi $ and other particles are the elements of the exterior algebra defined by this space. These vectors are the Grassmann numbers, they indeed define a basis for the exterior powers of the $\chi$ (wedging kind of replaced by the Clifford product).
\subsection{Algebra for the Standard Model}
We construct an algebra  over the space $\chi^{\dagger} \oplus \chi $ and generate a basis of null vectors \cite{Stoica_2018}. The two chiral spaces are the maximally isotropic subspaces for the inner product. So from our previous knowledge and definitions in the earlier section, we have the following:-
\begin{equation}
    \chi = \bigl\{q_{1}, q_{2} , q_{3} \bigl\}  \:\:\:\:\:\:\:\:\:\:\: 
\chi^{\dagger} = \bigl\{q_{1}^{\dagger} ,q_{2}^{\dagger} , q_{3}^{\dagger} \bigl\}
\end{equation}
\begin{equation}
    \bigl\{q_{i}^{\dagger}, q_{j}^{\dagger} \bigl\} = 0 \:\:\:\:\:\:\:\:\:\:\:\:\:\:\:\:\:
\bigl\{q_{i}, q_{j} \bigl\} = 0 \:\:\:\:\:\:\:\:\:\:\:\:\:\:\:\:\: \bigl\{q_{i}, q_{j}^{\dagger} \bigl\} = \delta_{ij}
\end{equation}
\begin{equation}
    q = q_{1}q_{2}q_{3} \:\:\:\:\:\:\:\:\:\:\:\:\:\:\:\:\:
q^{\dagger} =q_{3}^{\dagger}q_{2}^{\dagger}q_{1}^{\dagger}
\end{equation}
\begin{equation}
    p = qq^{\dagger} \:\:\:\:\:\:\:\:\:\:\:\:\:\:\:\:\:
p^{\prime} = q^{\dagger}q
\end{equation}
Here 
$p$ and $p^{\prime}$ are the idempotents; $q$ and $q^{\dagger}$ are the nilpotents as defined earlier. 
 We can now define an orthonormal basis using these null vectors, by the following construction:-
 \begin{equation}
     \chi^{\dagger} \oplus \chi = \bigl\{e_{1}, e_{2} , e_{3}, \Tilde{e}_{1},\Tilde{e}_{2},\Tilde{e}_{3} \bigl\}
 \end{equation}
 \begin{equation}
     e_{j} = q_{j} + q_{j}^{\dagger}
 \end{equation}
 \begin{equation}
     \Tilde{e}_{j} = i(q_{j}^{\dagger} - q_{j} )
 \end{equation}
 \begin{equation}
     e = e_{1}e_{2}e_{3} \:\:\:\:\:\:\:\:\:\:\:\:\:\:\:\:\: \Tilde{e} = \Tilde{e}_{1}\Tilde{e}_{2}\Tilde{e}_{3}
 \end{equation}
 \begin{equation}
     e_{i}^{2} = \Tilde{e}_{i}^{2} =1
 \end{equation}
 \begin{equation}
     e\Tilde{e} = -\Tilde{e}e
 \end{equation}
 Observe that we could have chosen $-\Tilde{e}_{j}$ as the orthonormal vector instead of $e_{j}$, this will change the definition of null vectors in terms of the orthonormal vectors. Here however we choose the above given definitions.
\subsection{Ideals and Representations}
We recall that
\begin{equation}
    Cl^{even}(7) \cong Cl(6) \cong M_{8}(\mathbb{C})
\end{equation}
We know that the $Cl(7)$ spinors have representations as the elements of the $Cl(6)$ algebra. We construct left ideals in the $Cl(6)$ algebra and now left multiply various elements of the $Cl(6)$ algebra with the elements of the left ideal; as the space is closed, the resulting space is invariant. It gives us the matrix representations of the elements of $Cl(6)$. Following the earlier framework \cite{furey2016standard}, we will act with the creation operators on the idempotents to create the particles and thus get the representation of particles in the algebra. A basis of the minimal left ideal or the action of all creation operators on one idempotent can be written as \cite{Stoica_2018}:-
\begin{equation}
    \bigl\{p, q_{23}^{\dagger}p, q_{31}^{\dagger}p, q_{12}^{\dagger}p, q_{321}^{\dagger}p,q_{1}^{\dagger}p, q_{2}^{\dagger}p, q_{3}^{\dagger}p \bigl\}
\end{equation}
On the simplification of the above given basis in terms of the $q_{i}$'s we have:-
\begin{equation}
    \bigl\{qq^{\dagger} , -q_{1}q^{\dagger}, -q_{2}q^{\dagger} , -q_{3}q^{\dagger}, q^{\dagger}, q_{23}q^{\dagger} , q_{31}q^{\dagger} , q_{12}q^{\dagger}\bigl\}
\end{equation}
We left act on this algebraic basis using various creation and annihilation operators. It will give us the representations of the algebra as the endomorphisms on the underlying vector space. For the algebraic ideal $A$, we have:-
\begin{equation}
    \rho: A \longrightarrow \rho(A)
\end{equation}
\begin{equation}
    \rho(A) : b \in A \longrightarrow \rho(A)(b) \in A
\end{equation}
\begin{equation}
    \rho(A) \cong End_{C}(A \cong_{vec} C^{8})
\end{equation}\\
Using the above information we have :-
\begin{equation}
    \begin{bmatrix}
qq^{\dagger} \\ -q_{1}q^{\dagger}\\ -q_{2}q^{\dagger} \\ -q_{3}q^{\dagger}\\ q^{\dagger}\\ q_{23}q^{\dagger} \\ q_{31}q^{\dagger} \\ q_{12}q^{\dagger}
\end{bmatrix} \overset{q_{1}^{\dagger}}{\longrightarrow}
\begin{bmatrix}
0\\0\\-q_{1}q_{2}q^{\dagger}\\-q_{1}q_{3}q^{\dagger}\\q_{1}q^{\dagger} \\ qq^{\dagger}\\0\\0
\end{bmatrix}
\end{equation}
So the action of $q^{\dagger}_{1}$ can be reprsented as :-
\begin{equation}
    \begin{bmatrix}
0\\0\\-q_{1}q_{2}q^{\dagger}\\-q_{1}q_{3}q^{\dagger}\\q_{1}q^{\dagger} \\ qq^{\dagger}\\0\\0
\end{bmatrix} = 
\begin{bmatrix}
0 & 0 & 0 & 0 & 0 & 0 & 0 & 0\\
0 & 0 & 0 & 0 & 0 & 0 & 0 & 0\\
0 & 0 & 0 & 0 & 0 & 0 & 0 & -1\\
0 & 0 & 0 & 0 & 0 & 0 & 1 & 0\\
0 & -1 & 0 & 0 & 0 & 0 & 0 & 0\\
1 & 0 & 0 & 0 & 0 & 0 & 0 & 0\\
0 & 0 & 0 & 0 & 0 & 0 & 0 & 0\\
0 & 0 & 0 & 0 & 0 & 0 & 0 & 0\\
\end{bmatrix}
\begin{bmatrix}
qq^{\dagger} \\ -q_{1}q^{\dagger}\\ -q_{2}q^{\dagger} \\ -q_{3}q^{\dagger}\\ q^{\dagger}\\ q_{23}q^{\dagger} \\ q_{31}q^{\dagger} \\ q_{12}q^{\dagger}
\end{bmatrix}
\end{equation}
Matrix reprsentation of the $q_{1}^{\dagger}$ and other null vectors is therefore given below:-
\begin{equation}
    q_{1}^{\dagger} = \begin{bmatrix}
0 & 0 & 0 & 0\\
0 & 0 & 0 & -i\sigma_{2}\\
-i\sigma_{2} & 0 & 0 & 0\\
0 & 0 & 0 & 0\\
\end{bmatrix} \:\:\:\ q_{2}^{\dagger} =  \begin{bmatrix}
0 & 0 & 0 & \sigma_{3}^{-}\\
0 & 0 & -\sigma_{3}^{-} & 0\\
0 & -\sigma_{3}^{+} & 0 & 0\\
\sigma_{3}^{+} & 0 & 0 & 0 
\end{bmatrix}
\end{equation}
\begin{equation}
    q_{3}^{\dagger} =  \begin{bmatrix}
0 & 0 & 0 & -\sigma_{-}\\
0 & 0 & \sigma_{+} & 0\\
0 & -\sigma_{+} & 0 & 0\\
\sigma_{-} & 0 & 0 & 0 
\end{bmatrix}
\end{equation}
\begin{equation}
    \sigma_{+} = \begin{bmatrix}
0 & 1\\
0 & 0
\end{bmatrix} \:\:\:\:\
\sigma_{-} = \begin{bmatrix}
0 & 0\\
1 & 0
\end{bmatrix} \:\:\:\:\
\sigma_{3}^{+} =  \begin{bmatrix}
1 & 0\\
0 & 0
\end{bmatrix}  \:\:\:\:\
\sigma_{3}^{-} =  \begin{bmatrix}
0 & 0\\
0 & 1
\end{bmatrix}
\end{equation}
With the matrix definitions of the null vectors we have matrix representations for other defined elements as given below, the nilpotents, idempotents and the orthonormal vectors respectively. 
\begin{equation}
    q^{\dagger}=
\begin{bmatrix}
0 & 0 & 0 & 0\\
0 & 0 & 0 & 0\\
\sigma_{3}^{+} & 0 & 0 & 0\\
0 & 0 & 0 & 0 
\end{bmatrix}  \:\:\:\:\:\:\:\:\:\:\:\:\:\
q =
\begin{bmatrix}
0 & 0 & \sigma_{3}^{+} & 0\\
0 & 0 & 0 & 0\\
0 & 0 & 0 & 0\\
0 & 0 & 0 & 0 
\end{bmatrix}
\end{equation}
\begin{equation}
    p =
\begin{bmatrix}
\sigma_{3}^{+} & 0 & 0 & 0\\
0 & 0 & 0 & 0\\
0 & 0 & 0 & 0\\
0 & 0 & 0 & 0 
\end{bmatrix} \:\:\:\:\:\:\:\:\:\:\:\:\:\ 
p^{\prime} =
\begin{bmatrix}
0 & 0 & 0 & 0\\
0 & 0 & 0 & 0\\
0 & 0 & \sigma_{3}^{+} & 0\\
0 & 0 & 0 & 0 
\end{bmatrix}
\end{equation}
The orthonormal vectors are given below :-
\begin{equation}
    e_{1}= 
\begin{bmatrix}
0 & 0 & i\sigma_{2} & 0\\
0 & 0 & 0 & -i\sigma_{2}\\
-i\sigma_{2} & 0 & 0 & 0\\
0 & i\sigma_{2} & 0 & 0 
\end{bmatrix} \:\:\:\:\:\:\:\:\:\:\:\:\:\:\:\:\:\:\:\:\:\:\
e_{2} = 
\begin{bmatrix}
0 & 0 & 0 & 1_{2}\\
0 & 0 & -1_{2} & 0\\
0 & -1_{2} & 0 & 0\\
1_{2} & 0 & 0 & 0
\end{bmatrix}
\end{equation}
\begin{equation}
    e_{3} =
\begin{bmatrix}
0 & 0 & 0 & i\sigma_{2} \\
0 & 0 & i\sigma_{2} & 0\\
0 & -i\sigma_{2} & 0 & 0\\
-i\sigma_{2} & 0 & 0 & 0
\end{bmatrix} \:\:\:\:\:\:\:\:\:\:\:\:\:\:\:\:\:\:\:\:\:\:\
\Tilde{e}_{1} = 
\begin{bmatrix}
0 & 0 & \sigma_{2} & 0\\
0 & 0 & 0 & \sigma_{2}\\
\sigma_{2} & 0 & 0 & 0\\
0 & \sigma_{2} & 0 & 0 
\end{bmatrix}
\end{equation}
\begin{equation}
    \Tilde{e}_{2} =
\begin{bmatrix}
0 & 0 & 0 & -i\sigma_{3} \\
0 & 0 & i\sigma_{3} & 0\\
0 & -i\sigma_{3} & 0 & 0\\
i\sigma_{3} & 0 & 0 & 0
\end{bmatrix} \:\:\:\:\:\:\:\:\:\:\:\:\:\
\Tilde{e}_{3} =
\begin{bmatrix}
0 & 0 & 0 & -i\sigma_{1} \\
0 & 0 & i\sigma_{1} & 0\\
0 & -i\sigma_{1} & 0 & 0\\
i\sigma_{1} & 0 & 0 & 0
\end{bmatrix}
\end{equation}
To compute the inner product between various orthonormal vectors use the matrix multiplication :-
\begin{equation}
    \Vec{a}.\Vec{b} = \frac{1}{2}(ab + ba)
\end{equation}
\subsubsection{SU(2) Symmetry}
We will first partition this 8-dimensional space into a vector sum of two irreducible spaces of dimension 4. Then these 4 dimensional spaces have to be further decomposed into irreducible subspaces, defined to be of different chirality. To proceed, we need to define new matrix operators; for the weak isospin $\frac{1}{2}$ and $-\frac{1}{2}$, we use an isospin operator (it decomposes the space into two irreducible representations).
\begin{equation}
    e=
\begin{bmatrix}
0 & 1_{4}\\
-1_{4} & 0
\end{bmatrix} \:\:\:\:\:\:\:\:\:\:\:\:\:\:\:\
\Tilde{e} = i
\begin{bmatrix}
0 & 1_{4}\\
1_{4} & 0
\end{bmatrix}
\end{equation}
\begin{equation}
    e\Tilde{e} = i
\begin{bmatrix}
1_{4} & 0\\
0 & -1_{4}
\end{bmatrix}
\end{equation} 
The $\frac{1 \pm e\Tilde{e}}{2}$ operator partitions the $\mathbb{C}^8$ space into two $\mathbb{C}^4$ spaces. We have the chirality operator given below :-
\begin{equation}
    \Gamma^{5} = -ie_{1}\Tilde{e}_{1} = 
\begin{bmatrix}
1_{2} & 0 & 0 & 0\\
0 & -1_{2} & 0 & 0\\
0 & 0 & -1_{2} & 0\\
0 & 0 & 0 & 1_{2}
\end{bmatrix}
\end{equation}
This operator can be used to define projectors on left and right chiral subspaces of two irreducible representations. Minus sign of the chirality operator represents the left chiral subspace. We need to mix the left chiral subspace of the particles for a given $SU_{L}(2)$ doublet. 
We can define a new basis of null vectors for the excited weak iso-spin states  as given below \cite{Stoica_2018}:-
\begin{equation}
    w_{u} = \begin{bmatrix}
0 & 1_{2} & 0 & 0\\
0 & 0 & 0 & 0\\
0 & 0  & 0 & -1_{2}\\
0 & 0 & 0 & 0
\end{bmatrix} \:\:\:\:\:\:\:\:\:\:\:\:\:\:\:\:\:\:\:\:\
w_{d} = \begin{bmatrix}
0 & 0 & -1_{2} & 0\\
0 & 0 & 0 & 0\\
0 & 0  & 0 & -1_{2}\\
0 & 0 & 0 & 0
\end{bmatrix}
\end{equation}
\begin{equation}
    w_{o} = \begin{bmatrix}
\sigma_{+} & 0 & 0 & 0\\
0 & -\sigma_{+} & 0 & 0\\
0 & 0 & -\sigma_{+} & 0\\
0 & 0 & 0 & \sigma_{+} 
        \end{bmatrix}
\end{equation}
\begin{equation}
    \bigl\{w_{i}, w_{j} \bigl\} = 0 \:\:\:\:\:\:\:\:\:\:\:\:\:\:\:\:\:\:\:\:\  \bigl\{w_{i}^{\dagger}, w_{j}^{\dagger} \bigl\}  = 0 \:\:\:\:\:\:\:\:\:\:\:\:\:\:\:\:\:\:\:\:\  \bigl\{w_{i}, w_{j}^{\dagger} \bigl\} = \delta_{ij}
\end{equation}
We have the matrix representations of the various elements; we can make the following identifications:-
\begin{equation}
\begin{bmatrix}
    p \\ w_{o}^{\dagger}p \\ w_{u}^{\dagger}p \\ w_{u}^{\dagger}w_{o}^{\dagger}p \\ w_{d}^{\dagger}p \\ w_{d}^{\dagger}w_{o}^{\dagger}p \\ w_{d}^{\dagger}w_{u}^{\dagger}p \\ w_{d}^{\dagger}w_{u}^{\dagger}w_{o}^{\dagger}p \\
\end{bmatrix}=
    \begin{bmatrix}
        p\\ q_{23}^{\dagger}p\\ q_{31}^{\dagger}p\\ q_{12}^{\dagger}p\\ q_{321}^{\dagger}p\\q_{1}^{\dagger}p\\ q_{2}^{\dagger}p\\ q_{3}^{\dagger}p 
    \end{bmatrix}
\end{equation}
$w_{u}^{\dagger}$ represents the creation of left chiral subspace for an up-isospin particle from the idempotent, similarly $w_{d}^{\dagger}$ represents a creation of left chiral subspace for a down-isospin particle. $w_{d}^{\dagger}w_{u}^{\dagger}$ represents a creation of right chiral subspace of a down-isospin particle \cite{Stoica_2018}.
Observe the following decomposition, due to isospin projectors and later projections due to the chirality operator. $W_{C^{j}}$ represents  $j$ dimensional complex space.
\begin{equation}
    W_{\mathbb{C}^{8}} = W^{1}_{\mathbb{C}^{4}} \oplus  W^{2}_{\mathbb{C}^{4}}
\end{equation}
\begin{equation}
    W_{\mathbb{C}^{8}} = W^{1}_{\mathbb{C}^{2},R} \oplus  W^{1}_{\mathbb{C}^{2},L} \oplus W^{2}_{\mathbb{C}^{2},L} \oplus  W^{2}_{\mathbb{C}^{2},R}
\end{equation}
\begin{equation}
    =\:\:\:  \{ p, w_{o}^{\dagger}p \} \oplus \{ w_{u}^{\dagger}p , w_{u}^{\dagger}w_{o}^{\dagger}p \} \oplus \{ w_{d}^{\dagger}p, w_{d}^{\dagger}w_{o}^{\dagger}p \} \oplus \{w_{d}^{\dagger}w_{u}^{\dagger}p, w_{d}^{\dagger}w_{u}^{\dagger}w_{o}^{\dagger}p \}
\end{equation}
\begin{equation}
    = \:\:\:  \{ p, q_{23}^{\dagger}p \} \oplus \{q_{31}^{\dagger}p, q_{12}^{\dagger}p \} \oplus \{ q_{321}^{\dagger}p, q_{1}^{\dagger}p \} \oplus \{q_{2}^{\dagger}p, q_{3}^{\dagger}p \}
\end{equation}
We now define $SU(2)$ symmetry generators, these will only mix the left  chiral space for both fermions.
\begin{equation}
T_{1}= \frac{1}{2} \begin{pmatrix}
        0 & 0 & 0 & 0\\
        0 & 0 & 1_{2} & 0\\
        0 & 1_{2} & 0 & 0\\
        0 & 0 & 0 & 0
    \end{pmatrix}\:\:\:\:\:\:\:\:\:\:\:\:\:\:\:\:\:\:\:\:\
    T_{2} = \frac{-i}{2} \begin{pmatrix}
        0 & 0 & 0 & 0\\
        0 & 0 & -1_{2} & 0\\
        0 & 1_{2} & 0 & 0\\
        0 & 0 & 0 & 0
    \end{pmatrix}
\end{equation}
\begin{equation}
    T_{3} = \frac{1}{2} \begin{pmatrix}
        0 & 0 & 0 & 0\\
        0 & -1_{2} & 0 & 0\\
        0 & 0 & 1_{2} & 0\\
        0 & 0 & 0 & 0
    \end{pmatrix}
\end{equation}
\begin{equation}
    [T_{i},T_{j}] = i \epsilon_{ijk} T_{k}
\end{equation}
Observe that no mixing takes place for the right chiral space.
\subsubsection{Complete Space of Ideals}
The complete basis of the algebra  $Cl(6)$ in terms of the minimal left ideal can be written as given below, the initial basis is expanded via the right multiplication on that ideal.
\begin{equation}
    \begin{bmatrix}
p & pq_{23} & pq_{31}  & pq_{12} & pq_{321} & pq_{1} & pq_{2} & pq_{3} \\
q^{\dagger}_{23}p & q^{\dagger}_{23}pq_{23} & q^{\dagger}_{23}pq_{31}  & q^{\dagger}_{23}pq_{12} & q^{\dagger}_{23}pq_{321} & q^{\dagger}_{23}pq_{1} & q^{\dagger}_{23}pq_{2} & q^{\dagger}_{23}pq_{3} \\
q^{\dagger}_{31}p & q^{\dagger}_{31}pq_{23} & q^{\dagger}_{31}pq_{31}  & q^{\dagger}_{31}pq_{12} & q^{\dagger}_{31}pq_{321} & q^{\dagger}_{31}pq_{1} & q^{\dagger}_{31}pq_{2} & q^{\dagger}_{31}pq_{3}\\
q^{\dagger}_{12}p & q^{\dagger}_{12}pq_{23} & q^{\dagger}_{12}pq_{31}  & q^{\dagger}_{12}pq_{12} & q^{\dagger}_{12}pq_{321} & q^{\dagger}_{12}pq_{1} & q^{\dagger}_{12}pq_{2} & q^{\dagger}_{12}pq_{3}\\ 
q^{\dagger}_{321}p & q^{\dagger}_{321}pq_{23} & q^{\dagger}_{321}pq_{31}  & q^{\dagger}_{321}pq_{12} & q^{\dagger}_{321}pq_{321} & q^{\dagger}_{321}pq_{1} & q^{\dagger}_{321}pq_{2} & q^{\dagger}_{321}pq_{3} \\
q_{1}^{\dagger}p & q_{1}^{\dagger}pq_{23} & q_{1}^{\dagger}pq_{31}  & q_{1}^{\dagger}pq_{12} & q_{1}^{\dagger}pq_{321} & q_{1}^{\dagger}pq_{1} & q_{1}^{\dagger}pq_{2} & q_{1}^{\dagger}pq_{3}\\
q_{2}^{\dagger}p & q_{2}^{\dagger}pq_{23} & q_{2}^{\dagger}pq_{31}  & q_{2}^{\dagger}pq_{12} & q_{2}^{\dagger}pq_{321} & q_{2}^{\dagger}pq_{1} & q_{2}^{\dagger}pq_{2} & q_{2}^{\dagger}pq_{3} \\
q_{3}^{\dagger}p & q_{3}^{\dagger}pq_{23} & q_{3}^{\dagger}pq_{31}  & q_{3}^{\dagger}pq_{12} & q_{3}^{\dagger}pq_{321} & q_{3}^{\dagger}pq_{1} & q_{3}^{\dagger}pq_{2} & q_{3}^{\dagger}pq_{3}
\end{bmatrix}_{8 \times 8}
\end{equation}
Now we can identify 4-dimensional spaces using the classifier spaces, isospin spaces and spinor chiral spaces with various particles \cite{gillard2019cell8}. We use the elements from row 1, and row 5 to assign the electric charge  to the two 4-dimensional column spinors present in a column by calculating the total electric charge from the product of the creation and the annihilation operators. We identify these basis with the following particle spaces, remember that these are now complex numbers on which $M_{8}(C)$ can act from the left.
\begin{equation}
    \begin{bmatrix}
\nu_{R_{1}} & u^{r}_{R_{1}} & u^{b}_{R_{1}} & u^{g}_{R_{1}} & \overline{e}_{L_{1}} & \overline{d}^{\overline{r}}_{L_{1}} & \overline{d}^{\overline{b}}_{L_{1}} & \overline{d}^{\overline{g}}_{L_{1}}\\
\nu_{R_{2}} & u^{r}_{R_{2}} & u^{b}_{R_{2}} & u^{g}_{R_{2}} & \overline{e}_{L_{2}} & \overline{d}^{\overline{r}}_{L_{2}} & \overline{d}^{\overline{b}}_{L_{2}} & \overline{d}^{\overline{g}}_{L_{2}}\\
\nu_{L_{1}} & u^{r}_{L_{1}} & u^{b}_{L_{1}} & u^{g}_{L_{1}} & \overline{e}_{R_{1}} & \overline{d}^{\overline{r}}_{R_{1}} & \overline{d}^{\overline{b}}_{R_{1}} & \overline{d}^{\overline{g}}_{R_{1}}\\
\nu_{L_{2}} & u^{r}_{L_{2}} & u^{b}_{L_{2}} & u^{g}_{L_{2}} & \overline{e}_{R_{2}} & \overline{d}^{\overline{r}}_{R_{2}} & \overline{d}^{\overline{b}}_{R_{2}} & \overline{d}^{\overline{g}}_{R_{2}}\\
e_{L_{1}} & d_{L_{1}}^{r} & d_{L_{1}}^{g} & d_{L_{1}}^{b} & \overline{\nu}_{R_{1}} & \overline{u}_{R_{1}}^{\overline{r}} & \overline{u}_{R_{1}}^{\overline{b}} & \overline{u}_{R_{1}}^{\overline{g}}\\
e_{L_{2}} & d_{L_{2}}^{r} & d_{L_{2}}^{g} & d_{L_{2}}^{b} & \overline{\nu}_{R_{2}} & \overline{u}_{R_{2}}^{\overline{r}} & \overline{u}_{R_{2}}^{\overline{b}} & \overline{u}_{R_{2}}^{\overline{g}}\\
e_{R_{1}} & d_{R_{1}}^{r} & d_{R_{1}}^{g} & d_{R_{1}}^{b} & \overline{\nu}_{L_{1}} & \overline{u}_{L_{1}}^{\overline{r}} & \overline{u}_{L_{1}}^{\overline{b}} & \overline{u}_{L_{1}}^{\overline{g}}\\
e_{R_{2}} & d_{R_{2}}^{r} & d_{R_{2}}^{g} & d_{R_{2}}^{b} & \overline{\nu}_{L_{2}} & \overline{u}_{L_{2}}^{\overline{r}} & \overline{u}_{L_{2}}^{\overline{b}} & \overline{u}_{L_{2}}^{\overline{g}}\\
\end{bmatrix}_{Charge}
\end{equation}

\subsubsection{Left Action on the Space of Ideals}
Now we have arranged our total complex ideal space in such a manner that left multiplication will only cause transformation within an ideal. We have already shown our $SU(2)$ generators and their intended action on an ideal (a $C^{8}$ column, basically). It is important to notice that $Cl(6)\cong Cl(4) \otimes Cl(2)$, now $Cl(4)$ represents the Dirac algebra and $Cl(2)$ represents the spin algebra. Essential transformations will basically be Lorentzian in nature and $SU(2)$ transformations. Thus if we want to include spin in our analysis, we can do so by looking at the algebra $Cl(4)\otimes Cl(2)\otimes Cl(2) \cong Cl(4)_{Dirac}\otimes Cl(2)_{Iso-spin} \otimes Cl(2)_{Spin}$ and the left action of various elements of $Cl(8)$ algebra on the ideals of the $Cl(8)$.
\subsubsection{Right Action on the Space of Ideals}
Looking at the total space of ideals, we see that a right multiplication by $M_{8}(C)$ will permute the columns. It can basically change the colour space of various quarks. So here, essential transformations for us will be $SU(3) $  transformations. The matrices that can do so will form one-to-one correspondence with Gell-Mann's $SU(3)$ matrices \cite{gillard2019cell8}.
\subsection{Cl(7) Algebra}
We have :-
\begin{equation}
    Cl(7) = \mathbb{C} \times \mathbb{O} \oplus \omega (\mathbb{C} \times \mathbb{O}) = Cl(6) \oplus Cl(6)
\end{equation}
With the above information, we proceed for the extended gauge group $SU(3)_{grav}\times SU(2)_{R} \times U_{g}(1)$.
\begin{table}[ht!]
\centering
    \begin{tabular}{||c c c c||} 
 \hline
 Force/Charge & Internal space & Dimension & Symmetry \\ [0.5ex] 
 \hline\hline
 Gravi-Electromagnetism & $\chi_{gem}$ & 1 & U(1) \\ 
 \hline
 Gravi-Strong & $\chi_{grav}$ & 3 & SU(3) \\
 \hline
 Gravi-Weak Hypercharge & $\chi_{g}$ & 1 & U(1) \\
 \hline
 Gravi Weak- Electromagnetism & $\chi_{g-ew}$ & 2 & U(2) \\ [1ex] 
 \hline
\end{tabular}
\caption{New Symmetry Group}
\end{table}
 With this we can define a new internal space, as done earlier for all the particles. It is done in Table $6$.
\begin{table}[ht!]
\centering
    \begin{tabular}{||c  c||} 
 \hline
 Particle & Internal Space \\ [0.5ex] 
 \hline\hline
 d & $\chi_{gem}^{3}$ \\ 
 \hline
 $\overline{u}$ & $\chi_{gem}^{2}\overline{\chi}_{grav}$ \\
 \hline
 $e^{-}$ & $\chi_{gem}\chi_{grav}$ \\
 \hline
 $\overline{\nu}$ & $\mathbb{C}$ \\
 \hline
 $\nu$ & $\mathbb{C}$  \\
 \hline
 $e^{+}$ & $\chi_{gem}^{-1}\overline{\chi}_{grav}$ \\
 \hline
 u & $\chi_{gem}^{-2}\chi_{grav}$ \\
 \hline 
 $\overline{d}$ & $\chi_{gem}^{-3}$ \\ [1ex] 
 \hline
\end{tabular}
\caption{Internal Space due to extended Symmetry Group}
\end{table}\\
As done earlier, we will again define a space $\chi$ as done below:-
\begin{equation}
    \chi \cong \overline{\chi}_{gem}\otimes \overline{\chi}_{grav}
\end{equation}
The space $\chi_{gem}$ assigns $-\frac{1}{3}$ units of the gravi-charge to the particles.
We again have three null basis vectors for this tensor product space. Every basis represents a \textbf{gravi charge} of $\frac{1}{3}$ and each three anti-colour related to $SU(3)_{grav}$. \textbf{gravi-charge} is additive in nature and it will add up for a product of the null basis vectors. For $\chi$ space we denote the basis as : $\{ q_{i}^{\dagger} \}_{i= 1, 2,3}$ each basis vector has gravi-charge equal to $\frac{1}{3}$ and one gravi anti-colour. We then have the total space as $\chi \oplus \chi^{\dagger}$, with their basis vectors as given below:-
\begin{equation}
\chi = \bigl\{q_{1}^{\dagger} ,q_{2}^{\dagger} , q_{3}^{\dagger} \bigl\}\:\:\:\:\:\:\:\:\:\:\:\:\:\:\:\:\:\:\: 
   \chi^{\dagger} = \bigl\{q_{1}, q_{2} , q_{3} \bigl\}
\end{equation}
With this notation, we can proceed further and classify particles according to the representations of the exterior algebra. This has been done in Table $7$ and Table $8$.
\begin{table}[ht!]
\centering
    \begin{tabular}{||c  c||} 
 \hline
 Particle & Vectors in Exterior Space \\ [0.5ex] 
 \hline\hline
 $d$ & $\Lambda^{3}\overline{\chi}$ \\ 
 \hline
 $\overline{u}$ & $\Lambda^{2}\overline{\chi}$ \\
 \hline
 $e^{-}$ & $\Lambda^{1}\overline{\chi}$ \\
 \hline
 $\overline{\nu}$ & $\Lambda^{0}\overline{\chi}$ $\cong$ $\mathbb{C}$\\
 \hline
 $\nu$ & $\Lambda^{0}\chi$ $\cong$ $\mathbb{C}$ \\
 \hline
 $e^{+}$ & $\Lambda^{1}\chi$ \\
 \hline
 u & $\Lambda^{2}\chi$  \\
 \hline 
$ \overline{d}$ & $\Lambda^{3}\chi$  \\ [1ex] 
 \hline
\end{tabular}
\caption{Particles in Exterior Algebra}
\label{table:example_table}
\end{table}\\
Now for the other copy of the $Cl(6)$ we can use the complex conjugated vector space and similarly, $p^{\prime}$ as the idempotent. The new basis will then be:-
\begin{equation}
    \biggl\{ p^{\prime} , q_{23}p^{\prime} , q_{31}p^{\prime} , q_{12}p^{\prime} , q_{321}p^{\prime} , q_{1}p^{\prime},q_{2}p^{\prime},q_{3}p^{\prime}\biggl\}
\end{equation}
\begin{equation}
\biggl\{ q^{\dagger}q , q_{1}^{\dagger}q , q_{2}^{\dagger}q, q_{3}^{\dagger}q ,-q ,q^{\dagger}_{32}q , q_{13}^{\dagger}q , q_{21}^{\dagger}q \biggl\}
\end{equation}
Similarly, we can define the complete  space of ideals as defined earlier. 
\begin{equation}
    \begin{bmatrix}
p^{\prime} & p^{\prime}q^{\dagger}_{32} & p^{\prime}q^{\dagger}_{13}  & p^{\prime}q^{\dagger}_{21} & p^{\prime}q^{\dagger}_{321} & p^{\prime}q_{1}^{\dagger} & p^{\prime}q_{2}^{\dagger} & p^{\prime}q_{3}^{\dagger} \\
q_{23}p^{\prime} & q_{23}p^{\prime}q^{\dagger}_{32} & q_{23}p^{\prime}q^{\dagger}_{13}  & q_{23}p^{\prime}q^{\dagger}_{21} & q_{23}p^{\prime}q^{\dagger}_{321} & q_{23}p^{\prime}q_{1}^{\dagger} & q_{23}p^{\prime}q_{2}^{\dagger} & q_{23}p^{\prime}q_{3}^{\dagger} \\
q_{31}p^{\prime} & q_{31}p^{\prime}q^{\dagger}_{32} & q_{31}p^{\prime}q^{\dagger}_{13}  & q_{31}p^{\prime}q^{\dagger}_{21} & q_{31}p^{\prime}q^{\dagger}_{321} & q_{31}p^{\prime}q_{1}^{\dagger} & q_{31}p^{\prime}q_{2}^{\dagger} & q_{31}p^{\prime}q_{3}^{\dagger}\\
q_{12}p^{\prime} & q_{12}p^{\prime}q^{\dagger}_{32} & q_{12}p^{\prime}q^{\dagger}_{13}  & q_{12}p^{\prime}q^{\dagger}_{21} & q_{12}p^{\prime}q^{\dagger}_{321} & q_{12}p^{\prime}q_{1}^{\dagger} & q_{12}p^{\prime}q_{2}^{\dagger} & q_{12}p^{\prime}q_{3}^{\dagger} \\
q_{321}p^{\prime} & q_{321}p^{\prime}q^{\dagger}_{32} & q_{321}p^{\prime}q^{\dagger}_{13}  & q_{321}p^{\prime}q^{\dagger}_{21} & q_{321}p^{\prime}q^{\dagger}_{321} & q_{321}p^{\prime}q_{1}^{\dagger} & q_{321}p^{\prime}q_{2}^{\dagger} & q_{321}p^{\prime}q_{3}^{\dagger}\\
q_{1}p^{\prime} & q_{1}p^{\prime}q^{\dagger}_{32} & q_{1}p^{\prime}q^{\dagger}_{13}  & q_{1}p^{\prime}q^{\dagger}_{21} & q_{1}p^{\prime}q^{\dagger}_{321} & q_{1}p^{\prime}q_{1}^{\dagger} & q_{1}p^{\prime}q_{2}^{\dagger} & q_{1}p^{\prime}q_{3}^{\dagger} \\
q_{2}p^{\prime} & q_{2}p^{\prime}q^{\dagger}_{32} & q_{2}p^{\prime}q^{\dagger}_{13}  & q_{2}p^{\prime}q^{\dagger}_{21} & q_{2}p^{\prime}q^{\dagger}_{321} & q_{2}p^{\prime}q_{1}^{\dagger} & q_{2}p^{\prime}q_{2}^{\dagger} & q_{2}p^{\prime}q_{3}^{\dagger} \\
q_{3}p^{\prime} & q_{3}p^{\prime}q^{\dagger}_{32} & q_{3}p^{\prime}q^{\dagger}_{13}  & q_{3}p^{\prime}q^{\dagger}_{21} & q_{3}p^{\prime}q^{\dagger}_{321} & q_{3}p^{\prime}q_{1}^{\dagger} & q_{3}p^{\prime}q_{2}^{\dagger} & q_{3}p^{\prime}q_{3}^{\dagger}
\end{bmatrix}_{8 \times 8}
\end{equation}
As done earlier, we can get a matrix representation of the elements of $Cl(6)$ by the left action of various elements on the left ideal. 
We have a method to compute the $U(1)$ charges using the classifier space. We have employed this method to assign electric charges to 4 dimensional column vectors and hence classify the various subspaces of the complete space of ideals as particles. We use the same method and classify particles according to the \textbf{gravi-charges}.
\subsubsection{Right Adjoint Action}
The right action have a similar working. $M_{8}(C)$ acting from right can permute the columns and hence can cause colour changes for coloured particles. We have similar matrices for such transformation as we defined earlier for $SU(3)$. Here too, we can do the same for $SU(3)_{grav}$, the gravi-colour symmetry.
\subsubsection{Left Adjoint Action}
For the left action of the elements of the algebra, the space of ideals is closed. This gives us the matrix representations of the algebraic elements. But now we want our spinors such that they are $SU(2)_{R}$ active, it  means that their right chiral space mixes due to $SU(2)_{R}$. We can define a new basis of gravi weak isospin null vectors and similarly a set of $SU(2)$ generators. 
\begin{equation}
    \overline{\omega}_{u} =
\begin{bmatrix}
0 & -1_{2} & 0 & 0\\
0 & 0 & 0 & 0\\
0 & 0 & 0 & 1_{2}\\
0 & 0 & 0 & 0\\
\end{bmatrix} \:\:\:\:\:\:\:\:\:\:\:\:\:\:\:\:\:\:\:\:\:\:\:\:
\overline{\omega}_{d} = 
\begin{bmatrix}
0 & 0 & 0 & 0\\
0 & 0 & 0 & 0\\
-1_{2} & 0 & 0 & 0\\
0 & -1_{2} & 0 & 0\\
\end{bmatrix}
\end{equation}
\begin{equation}
    \\
\overline{\omega}_{o}=
-\begin{bmatrix}
\sigma_{+} & 0 & 0 & 0\\
0 & -\sigma_{+} & 0 & 0\\
0 & 0 & -\sigma_{+} & 0\\
0 & 0 & 0 & \sigma_{+}\\
\end{bmatrix}
\end{equation}
\begin{equation}
    \bigl\{\overline{w}_{i}, \overline{w}_{j} \bigl\} = 0 \:\:\:\:\:\:\:\:\:\:\:\:\:\:\:\:\:\:\:\:\  \bigl\{\overline{w}_{i}^{\dagger}, \overline{w}_{j}^{\dagger} \bigl\}  = 0 \:\:\:\:\:\:\:\:\:\:\:\:\:\:\:\:\:\:\:\:\  \bigl\{\overline{w}_{i}, \overline{w}_{j}^{\dagger} \bigl\} = \delta_{ij}
\end{equation}
\begin{eqnarray}
\begin{bmatrix}
p^{\prime} \\ q_{23}p^{\prime} \\ q_{31}p^{\prime} \\ q_{12}p^{\prime} \\q_{321}p^{\prime} \\ q_{1}p^{\prime}\\q_{2}p^{\prime}\\q_{3}p^{\prime}
\end{bmatrix}\longrightarrow
\begin{bmatrix}
p^{\prime} \\ \overline{\omega}_{o}^{\dagger} p^{\prime} \\ \overline{\omega_{u}}^{\dagger}p^{\prime} \\ \overline{\omega_{u}}^{\dagger}\overline{\omega_{o}}^{\dagger}p^{\prime} \\\overline{\omega_{d}}^{\dagger}p^{\prime} \\ \overline{\omega_{d}}^{\dagger}\overline{\omega_{o}}^{\dagger}p^{\prime}\\\overline{\omega}_{d}^{\dagger}\overline{\omega}_{u}^{\dagger}p^{\prime}\\\overline{\omega}_{d}^{\dagger}\overline{\omega}_{u}^{\dagger}\overline{\omega}_{o}^{\dagger}p^{\prime}
\end{bmatrix}\longrightarrow
\begin{bmatrix}
W_{1R}\\W_{1L}\\W_{2L}\\W_{2R}
\end{bmatrix}
\end{eqnarray}
Interpret these new null vectors as follows: $\overline{\omega}_{u}^{\dagger}$ as the creation operator of left chiral subspace of the gravi weak up isospin particle, $\overline{\omega}_{d}^{\dagger}$  as the creation operator to generate the left chiral subspace of the gravi weak down isospin particle. Similarly, $\overline{\omega}_{d}^{\dagger}\overline{\omega}_{u}^{\dagger}$ generates the right chiral subspace for gravi weak down isospin particle.
With these definitions for null basis we can define an orthonormal basis too, as defined earlier.
\begin{equation}
    \overline{u_{j}} = \overline{\omega}_{j} + \overline{\omega}_{j}^{\dagger}
\end{equation}
\begin{equation}
    \overline{u^{\prime}_{j}} = i(\overline{\omega}^{\dagger}_{j} - \overline{\omega}_{j})
\end{equation}
We have the following set of orthonormal vectors:-
\begin{equation}
    \biggl\{ \overline{u}_{u} , \overline{u}_{d} , \overline{u}_{o} , \overline{u^{\prime}}_{u} , \overline{u^{\prime}}_{d} ,  \overline{u^{\prime}}_{o}\biggl\}
\end{equation}
We now check the action of the SU(2) operator constructed from the $\overline{u}_{i}$ and $\overline{u}_{i}^{\prime}$. Define the new SU(2) generators as the following:-
\begin{eqnarray}
\overline{T}_{1} = \frac{1}{2}\begin{bmatrix}
0 & 0 & 0 & 1_{2}\\
0 & 0 & 0 & 0\\
0 & 0 & 0 & 0\\
1_{2} & 0 & 0 & 0
\end{bmatrix} \:\:\:\:\:\:\:\:\:\:\:\:\:\:\:\:\:\:\:\:\:\:\:\:
\overline{T}_{2} =  \frac{-i}{2}\begin{bmatrix}
0 & 0 & 0 & 1_{2}\\
0 & 0 & 0 & 0\\
0 & 0 & 0 & 0\\
-1_{2} & 0 & 0 & 0
\end{bmatrix}
\end{eqnarray}
\begin{equation}
\overline{T}_{3} = 
    \frac{1}{2}\begin{bmatrix}
1_{2} & 0 & 0 & 0\\
0 & 0 & 0 & 0\\
0 & 0 & 0 & 0\\
0 & 0 & 0 & -1_{2}
\end{bmatrix}
\end{equation}
\begin{equation}
    [\overline{T}_{i},\overline{T}_{j}] = i \epsilon_{ijk} \overline{T}_{k}
\end{equation}
Look carefully; it does not mix the left chiral components of the spinors from the two irreducible representations of different chirality. Hence it gives us  the gravitationally active right chiral spinors.
\subsubsection{Particle Identification}
Now we can proceed further and identify the various particles in the new complete space of ideals.
\begin{eqnarray}
\begin{bmatrix}
\nu_{R_{1}} & u^{r}_{R_{1}} & u^{b}_{R_{1}} & u^{g}_{R_{1}} & \overline{d}_{L_{1}} & \overline{e}^{\overline{r}}_{L_{1}} & \overline{e}^{\overline{b}}_{L_{1}} & \overline{e}^{\overline{g}}_{L_{1}}\\
\nu_{R_{2}} & u^{r}_{R_{2}} & u^{b}_{R_{2}} & u^{g}_{R_{2}} & \overline{d}_{L_{2}} & \overline{e}^{\overline{r}}_{L_{2}} & \overline{e}^{\overline{b}}_{L_{2}} & \overline{e}^{\overline{g}}_{L_{2}}\\
\nu_{L_{1}} & u^{r}_{L_{1}} & u^{b}_{L_{1}} & u^{g}_{L_{1}} & \overline{d}_{R_{1}} & \overline{e}^{\overline{r}}_{R_{1}} & \overline{e}^{\overline{b}}_{R_{1}} & \overline{e}^{\overline{g}}_{R_{1}}\\
\nu_{L_{2}} & u^{r}_{L_{2}} & u^{b}_{L_{2}} & u^{g}_{L_{2}} & \overline{d}_{R_{2}} & \overline{e}^{\overline{r}}_{R_{2}} & \overline{e}^{\overline{b}}_{R_{2}} & \overline{e}^{\overline{g}}_{R_{2}}\\
d_{L_{1}} & e_{L_{1}}^{r} & e_{L_{1}}^{g} & e_{L_{1}}^{b} & \overline{\nu}_{R_{1}} & \overline{u}_{R_{1}}^{\overline{r}} & \overline{u}_{R_{1}}^{\overline{b}} & \overline{u}_{R_{1}}^{\overline{g}}\\
d_{L_{2}} & e_{L_{2}}^{r} & e_{L_{2}}^{g} & e_{L_{2}}^{b} & \overline{\nu}_{R_{2}} & \overline{u}_{R_{2}}^{\overline{r}} & \overline{u}_{R_{2}}^{\overline{b}} & \overline{u}_{R_{2}}^{\overline{g}}\\
d_{R_{1}} & e_{R_{1}}^{r} & e_{R_{1}}^{g} & e_{R_{1}}^{b} & \overline{\nu}_{L_{1}} & \overline{u}_{L_{1}}^{\overline{r}} & \overline{u}_{L_{1}}^{\overline{b}} & \overline{u}_{L_{1}}^{\overline{g}}\\
d_{R_{2}} & e_{R_{2}}^{r} & e_{R_{2}}^{g} & e_{R_{2}}^{b} & \overline{\nu}_{L_{2}} & \overline{u}_{L_{2}}^{\overline{r}} & \overline{u}_{L_{2}}^{\overline{b}} & \overline{u}_{L_{2}}^{\overline{g}}\\
\end{bmatrix}_{Mass}
\end{eqnarray}
This has been done using the classifier space, weak force generators and $SU(3)$ operations. This will give us the following gravi-weak isospin doublets $SU(2)_{R}$ as given below.\\
First generation

\begin{eqnarray}
\begin{pmatrix}
u\\
e^{-}
\end{pmatrix}_{R} \:\:\:\:\:\:\:\:\:\:\:\:\:\:\:\:\:\: 
\begin{pmatrix}
\nu\\
d
\end{pmatrix}_{R}
\end{eqnarray}
Second Generation
\begin{eqnarray}
\begin{pmatrix}
t\\
\mu^{-}
\end{pmatrix}_{R} \:\:\:\:\:\:\:\:\:\:\:\:\:\:\:\:\:\: 
\begin{pmatrix}
\nu_{\mu}\\
b
\end{pmatrix}_{R}
\end{eqnarray}
Third Generation
\begin{eqnarray}
\begin{pmatrix}
c\\
\tau^{-}
\end{pmatrix}_{R} \:\:\:\:\:\:\:\:\:\:\:\:\:\:\:\:\:\: 
\begin{pmatrix}
\nu_{\tau}\\
s
\end{pmatrix}_{R}
\end{eqnarray}
\subsection{Triality and Cl(8) Algebra}
The basic reason to look into the $Cl(8)$ algebra is to use the \textbf{Triality} mapping. Triality mapping is, generally a very interesting object to study. Some authors have pointed towards its importance in studying three generations \cite{baez2002octonions, furey2016standard}.
\begin{equation}
    Cl^{even}(8) \cong Cl(7) \cong M_{8}(C) \oplus M_{8}(C)
\end{equation}
As explained earlier, $M_{8}(C) \oplus M_{8}(C)$ acts on a spinor space $S^{+}_{8}\oplus S^{-}_{8}$. Both $S^{+}_{8}$ and $S^{-}_{8}$ are 8-dimensional complex spinor spaces.
The eight generators of the $Cl(8)$ algebra give us the \textbf{vector} representation denoted by $V_{8}$. These can be considered as the basis vectors of the underlying vector space. Triality denoted by $t_{8}$ is defined as the following mapping \cite{baez2002octonions}:-
\begin{equation}
    t_{8} : S^{+}_{8} \times S^{-}_{8} \times V_{8} \longrightarrow \mathbb{C}
\end{equation}
So it basically takes three complex vector spaces and gives us a number as an output. Now focus on the space of the ideals for the $Cl(8)$ algebra. We have earlier seen that even subalgebra of $Cl(8)$ is same as $Cl(7)$ and we know that $Cl(7) \cong Cl(6) \oplus Cl(6)$, so the subspace - the even subalgebra of $Cl(8)$ is the same as the direct sum of the left ideal space of the two copies of $Cl(6)$.
\subsubsection{Space of Ideals in Cl(8)}
We require an 8-dimensional null basis to get the complete maximally totally isotropic subspace of the null vectors. To the 6-dimensional vector space of $\chi \oplus \chi^{\dagger}$ add a two-dimensional space $S$, for the two spin vectors. Our final underlying space will then be $\chi \oplus \chi^{\dagger} \oplus S $. To describe this new space, we also add $\{ q_{4} , q_{4}^{\dagger} \}$ to the pre-existing set of null vectors. Now any element in ideal will be a product from these 8 vectors, then we have \cite{gillard2019cell8}:-
\begin{equation}
    Cl(8) \cong Cl(4)\otimes Cl(2)\otimes Cl(2) \cong Cl(4)_{Dirac}\otimes Cl(2)_{Iso-spin} \otimes Cl(2)_{Spin}
\end{equation}
\begin{equation}
    \{ q_{1}, q_{2} ,q_{3} ,q_{4} ,q_{1}^{\dagger} ,q_{2}^{\dagger} ,q_{3}^{\dagger} ,q_{4}^{\dagger} \}
\end{equation}
\begin{equation}
    q = q_{1}q_{2}q_{3}q_{4} \:\:\:\:\:\:\:\:\:\:\:\:\:\:\:\:\:\:\  q^{\dagger} = q_{4}^{\dagger}q_{3}^{\dagger}q_{2}^{\dagger}q_{1}^{\dagger}
\end{equation}
\begin{equation}
     p = qq^{\dagger} \:\:\:\:\:\:\:\:\:\:\:\:\:\:\:\:\:\:\ p^{\prime} = q^{\dagger}q 
\end{equation}
Here $q$ and $q^{\dagger}$ are the nilpotents, the $p$ and $p^{\prime}$ are the idempotents. We use $p$ as the idempotent; from our previous information we know the importance of $Cl^{even}(8)$, so we can write the $Cl^{even}(8)$ ideal subspace as \cite{gillard2019cell8}:-

\begin{equation}
Cl(8)=
    \begin{bmatrix}
        Even_{1} & Odd\\
        Odd & Even_{2}
    \end{bmatrix}\implies
    Cl^{even}(8) = \begin{bmatrix}
        Even_{1} & 0\\
        0 & Even_{2}
    \end{bmatrix}
\end{equation}
The $Even_{1}$ part of the complete space of the ideal of $Cl(8)$ is given below:-
\begin{equation}
    \begin{bmatrix}
p & pq_{32} & pq_{13}  & pq_{21} & pq_{4321} & pq_{41} & pq_{42} & pq_{43} \\
q^{\dagger}_{23}p & q^{\dagger}_{23}pq_{32} & q^{\dagger}_{23}pq_{13}  & q^{\dagger}_{23}pq_{21} & q^{\dagger}_{23}pq_{4321} & q^{\dagger}_{23}pq_{41} & q^{\dagger}_{23}pq_{42} & q^{\dagger}_{23}pq_{43} \\
q^{\dagger}_{31}p & q^{\dagger}_{31}pq_{32} & q^{\dagger}_{31}pq_{13}  & q^{\dagger}_{31}pq_{21} & q^{\dagger}_{31}pq_{4321} & q^{\dagger}_{31}pq_{41} & q^{\dagger}_{31}pq_{42} & q^{\dagger}_{31}pq_{43}\\
q^{\dagger}_{12}p & q^{\dagger}_{12}pq_{32} & q^{\dagger}_{12}pq_{13}  & q^{\dagger}_{12}pq_{21} & q^{\dagger}_{12}pq_{4321} & q^{\dagger}_{12}pq_{41} & q^{\dagger}_{12}pq_{42} & q^{\dagger}_{12}pq_{43}\\ 
q^{\dagger}_{1234}p & q^{\dagger}_{1234}pq_{32} & q^{\dagger}_{1234}pq_{13}  & q^{\dagger}_{1234}pq_{21} & q^{\dagger}_{1234}pq_{4321} & q^{\dagger}_{1234}pq_{41} & q^{\dagger}_{1234}pq_{42} & q^{\dagger}_{1234}pq_{43}\\
q^{\dagger}_{14}p & q^{\dagger}_{14}pq_{32} & q^{\dagger}_{14}pq_{13}  & q^{\dagger}_{14}pq_{21} & q^{\dagger}_{14}pq_{4321} & q^{\dagger}_{14}pq_{41} & q^{\dagger}_{14}pq_{42} & q^{\dagger}_{14}pq_{43} \\
q^{\dagger}_{24}p & q^{\dagger}_{24}pq_{32} & q^{\dagger}_{24}pq_{13}  & q^{\dagger}_{24}pq_{21} & q^{\dagger}_{24}pq_{4321} & q^{\dagger}_{24}pq_{41} & q^{\dagger}_{24}pq_{42} & q^{\dagger}_{24}pq_{43} \\
q^{\dagger}_{34}p & q^{\dagger}_{34}pq_{32} & q^{\dagger}_{34}pq_{13}  & q^{\dagger}_{34}pq_{21} & q^{\dagger}_{34}pq_{4321} & q^{\dagger}_{34}pq_{41} & q^{\dagger}_{34}pq_{42} & q^{\dagger}_{34}pq_{43}
\end{bmatrix}
\end{equation}
The $Even_{2}$ part of the complete space of the ideal of $Cl(8)$ is given below:-
\begin{equation}
    \begin{bmatrix}
        q_{4}^{\dagger}pq_{4} & q_{4}^{\dagger}pq_{432} & q_{4}^{\dagger}pq_{413} & q_{4}^{\dagger}pq_{421} & q_{4}^{\dagger}pq_{321} &  q_{4}^{\dagger}pq_{1} & q_{4}^{\dagger}pq_{2} & q_{4}^{\dagger}pq_{3}\\
        q_{234}^{\dagger}pq_{4} & q_{234}^{\dagger}pq_{432} & q_{234}^{\dagger}pq_{413} & q_{234}^{\dagger}pq_{421} & q_{234}^{\dagger}pq_{321} &  q_{234}^{\dagger}pq_{1} & q_{234}^{\dagger}pq_{2} & q_{234}^{\dagger}pq_{3}\\
        q_{314}^{\dagger}pq_{4} & q_{314}^{\dagger}pq_{432} & q_{314}^{\dagger}pq_{413} & q_{314}^{\dagger}pq_{421} & q_{314}^{\dagger}pq_{321} &  q_{314}^{\dagger}pq_{1} & q_{314}^{\dagger}pq_{2} & q_{314}^{\dagger}pq_{3}\\
         q_{124}^{\dagger}pq_{4} & q_{124}^{\dagger}pq_{432} & q_{124}^{\dagger}pq_{413} & q_{124}^{\dagger}pq_{421} & q_{124}^{\dagger}pq_{321} &  q_{124}^{\dagger}pq_{1} & q_{124}^{\dagger}pq_{2} & q_{124}^{\dagger}pq_{3}\\
         q_{123}^{\dagger}pq_{4} & q_{123}^{\dagger}pq_{432} & q_{123}^{\dagger}pq_{413} & q_{123}^{\dagger}pq_{421} & q_{123}^{\dagger}pq_{321} &  q_{123}^{\dagger}pq_{1} & q_{123}^{\dagger}pq_{2} & q_{123}^{\dagger}pq_{3}\\
          q_{1}^{\dagger}pq_{4} & q_{1}^{\dagger}pq_{432} & q_{1}^{\dagger}pq_{413} & q_{1}^{\dagger}pq_{421} & q_{1}^{\dagger}pq_{321} &  q_{1}^{\dagger}pq_{1} & q_{1}^{\dagger}pq_{2} & q_{1}^{\dagger}pq_{3}\\
           q_{2}^{\dagger}pq_{4} & q_{2}^{\dagger}pq_{432} & q_{2}^{\dagger}pq_{413} & q_{2}^{\dagger}pq_{421} & q_{2}^{\dagger}pq_{321} &  q_{2}^{\dagger}pq_{1} & q_{2}^{\dagger}pq_{2} & q_{2}^{\dagger}pq_{3}\\
            q_{3}^{\dagger}pq_{4} & q_{3}^{\dagger}pq_{432} & q_{3}^{\dagger}pq_{413} & q_{3}^{\dagger}pq_{421} & q_{3}^{\dagger}pq_{321} &  q_{3}^{\dagger}pq_{1} & q_{3}^{\dagger}pq_{2} & q_{3}^{\dagger}pq_{3}\\
    \end{bmatrix}
\end{equation}
We know that there is a volume element in $Cl(7)$ algebra that can partition the algebra into two parts. Here the $Cl^{even}(8)$ gets partitioned into two parts depending upon whether the element is self-dual or not. So by this, we can assign different spins to both the even parts. Let us assign spin up to $Even_{1}$ and spin down to $Even_{2}$ part of the $Cl^{even}(8)$. By our previous arguments, we know that a correspondence can be established between each even part of the $Cl(8)$ algebra and two copies of $Cl(6)$, so we can identify a given subspace of even subalgebra by particles from one generation with two different definite spins.
Let us make some identifications; for example, for the $SU(2)_{L}$ active particles we can identify the $Even_{1}$ part as :-
\begin{equation}
A^{\uparrow} = 
    \begin{bmatrix}
\nu_{R_{1}}^{\uparrow} & u^{r \uparrow}_{R_{1}} & u^{b \uparrow}_{R_{1}} & u^{g \uparrow}_{R_{1}} & \overline{e}_{L_{1}}^{\uparrow} & \overline{d}^{\overline{r} \uparrow}_{L_{1}} & \overline{d}^{\overline{b} \uparrow}_{L_{1}} & \overline{d}^{\overline{g} \uparrow}_{L_{1}}\\
\nu_{R_{2}}^{\uparrow} & u^{r \uparrow}_{R_{2}} & u^{b \uparrow}_{R_{2}} & u^{g \uparrow}_{R_{2}} & \overline{e}_{L_{2}}^{\uparrow} & \overline{d}^{\overline{r} \uparrow}_{L_{2}} & \overline{d}^{\overline{b} \uparrow}_{L_{2}} & \overline{d}^{\overline{g} \uparrow}_{L_{2}}\\
\nu_{L_{1}}^{\uparrow} & u^{r \uparrow}_{L_{1}} & u^{b \uparrow}_{L_{1}} & u^{g \uparrow}_{L_{1}} & \overline{e}_{R_{1}}^{\uparrow} & \overline{d}^{\overline{r} \uparrow}_{R_{1}} & \overline{d}^{\overline{b} \uparrow}_{R_{1}} & \overline{d}^{\overline{g} \uparrow}_{R_{1}}\\
\nu_{L_{2}}^{\uparrow} & u^{r \uparrow}_{L_{2}} & u^{b \uparrow}_{L_{2}} & u^{g \uparrow}_{L_{2}} & \overline{e}_{R_{2}}^{\uparrow} & \overline{d}^{\overline{r} \uparrow}_{R_{2}} & \overline{d}^{\overline{b} \uparrow}_{R_{2}} & \overline{d}^{\overline{g} \uparrow}_{R_{2}}\\
e_{L_{1}}^{\uparrow} & d_{L_{1}}^{r \uparrow} & d_{L_{1}}^{g \uparrow} & d_{L_{1}}^{b \uparrow} & \overline{\nu}_{R_{1}}^{\uparrow} & \overline{u}_{R_{1}}^{\overline{r} \uparrow} & \overline{u}_{R_{1}}^{\overline{b} \uparrow} & \overline{u}_{R_{1}}^{\overline{g} \uparrow}\\
e_{L_{2}}^{\uparrow} & d_{L_{2}}^{r \uparrow} & d_{L_{2}}^{g \uparrow} & d_{L_{2}}^{b \uparrow} & \overline{\nu}_{R_{2}}^{\uparrow} & \overline{u}_{R_{2}}^{\overline{r} \uparrow} & \overline{u}_{R_{2}}^{\overline{b} \uparrow} & \overline{u}_{R_{2}}^{\overline{g} \uparrow}\\
e_{R_{1}}^{\uparrow} & d_{R_{1}}^{r \uparrow} & d_{R_{1}}^{g \uparrow} & d_{R_{1}}^{b \uparrow} & \overline{\nu}_{L_{1}}^{\uparrow} & \overline{u}_{L_{1}}^{\overline{r} \uparrow} & \overline{u}_{L_{1}}^{\overline{b} \uparrow} & \overline{u}_{L_{1}}^{\overline{g}\uparrow}\\
e_{R_{2}}^{\uparrow} & d_{R_{2}}^{r \uparrow} & d_{R_{2}}^{g \uparrow} & d_{R_{2}}^{b \uparrow} & \overline{\nu}_{L_{2}}^{\uparrow} & \overline{u}_{L_{2}}^{\overline{r} \uparrow} & \overline{u}_{L_{2}}^{\overline{b} \uparrow} & \overline{u}_{L_{2}}^{\overline{g} \uparrow}\\
\end{bmatrix}_{Charge}
\end{equation}
Similarly, the $Even_{2}$ part can be identified by the second generation $SU(2)_{L}$ particle eigenstates.
We replace the particles with the corresponding second generation particles.
\begin{equation}
    \{ \nu , \overline{\nu} \} \longrightarrow \{ \nu_{\mu} , \overline{\nu}_{\mu} \}
\end{equation}
\begin{equation}
    \{ e , \overline{e} \} \longrightarrow \{ \mu ,  \overline{\mu} \}
\end{equation}
\begin{equation}
    \{ u , d \} \longrightarrow \{ c , s\}
\end{equation}
\begin{equation}
    \{ \overline{u} , \overline{d} \} \longrightarrow \{ \overline{c} , \overline{s} \}
\end{equation}
However, this family will have opposite sign of spin, let us denote the second generation $SU(2)_{L}$ active family with down spin as $B^{\downarrow}$. Similarly, third generation family with up spin can be represented as $C^{\uparrow}$. So the total $SU(2)_{L}$ active vector spaces $(\mathbb{C}^{8}\times\mathbb{C}^{8})$ with different spins available to us can be listed below as following:-
\begin{equation}
    \{ A^{\uparrow} , A^{\downarrow} , B^{\uparrow} , B^{\downarrow} , C^{\uparrow} , C^{\downarrow} \}
\end{equation}
Now observe the following:-
\begin{equation}
    Cl(9) = Cl(7)\otimes Cl(2)  = (C \times O \oplus \omega(C \times O)) \otimes Cl(2) = Cl(8) \oplus Cl(8)
\end{equation}
Now we can use  one copy of $Cl(8)$ to construct the representations for left active $SU(2)_{L}$ particles. The other copy of $Cl(8)$ can be used to construct the right active $SU(2)_{R}$ particles. Both copies will give us the spin up and spin down particles. For $SU(2)_{R}$ active particles we can use the complexified space of ideals and use the $p^{\prime}$ as idempotent. We do the similar procedure, now again the $Cl^{even}(8)$ algebra will get partitioned into two subalgebras denoting different spins. An example of particle identification of different gravi-charges and $SU(2)_{R}$ active first generation is present below:-
\begin{eqnarray}
P^{\uparrow} = 
\begin{bmatrix}
\nu_{R_{1}}^{\uparrow} & u^{r \uparrow}_{R_{1}} & u^{b \uparrow}_{R_{1}} & u^{g \uparrow}_{R_{1}} & \overline{d}_{L_{1}} & \overline{e}^{\overline{r \uparrow}}_{L_{1}} & \overline{e}^{\overline{b} \uparrow}_{L_{1}} & \overline{e}^{\overline{g} \uparrow}_{L_{1}}\\
\nu_{R_{2}}^{\uparrow} & u^{r \uparrow}_{R_{2}} & u^{b \uparrow}_{R_{2}} & u^{g \uparrow}_{R_{2}} & \overline{d}_{L_{2}}^{\uparrow} & \overline{e}^{\overline{r} \uparrow}_{L_{2}} & \overline{e}^{\overline{b} \uparrow}_{L_{2}} & \overline{e}^{\overline{g} \uparrow}_{L_{2}}\\
\nu_{L_{1}}^{\uparrow} & u^{r \uparrow}_{L_{1}} & u^{b \uparrow}_{L_{1}} & u^{g \uparrow}_{L_{1}} & \overline{d}_{R_{1}}^{\uparrow} & \overline{e}^{\overline{r} \uparrow}_{R_{1}} & \overline{e}^{\overline{b} \uparrow}_{R_{1}} & \overline{e}^{\overline{g}\uparrow}_{R_{1}}\\
\nu_{L_{2}}^{\uparrow} & u^{r \uparrow}_{L_{2}} & u^{b \uparrow}_{L_{2}} & u^{g \uparrow}_{L_{2}} & \overline{d}_{R_{2}}^{\uparrow} & \overline{e}^{\overline{r} \uparrow}_{R_{2}} & \overline{e}^{\overline{b} \uparrow}_{R_{2}} & \overline{e}^{\overline{g} \uparrow}_{R_{2}}\\
d_{L_{1}}^{\uparrow} & e_{L_{1}}^{r \uparrow} & e_{L_{1}}^{g \uparrow} & e_{L_{1}}^{b \uparrow} & \overline{\nu}_{R_{1}}^{\uparrow} & \overline{u}_{R_{1}}^{\overline{r} \uparrow} & \overline{u}_{R_{1}}^{\overline{b} \uparrow} & \overline{u}_{R_{1}}^{\overline{g} \uparrow}\\
d_{L_{2}}^{\uparrow} & e_{L_{2}}^{r \uparrow} & e_{L_{2}}^{g \uparrow} & e_{L_{2}}^{b \uparrow} & \overline{\nu}_{R_{2}}^{\uparrow} & \overline{u}_{R_{2}}^{\overline{r} \uparrow} & \overline{u}_{R_{2}}^{\overline{b} \uparrow} & \overline{u}_{R_{2}}^{\overline{g} \uparrow}\\
d_{R_{1}}^{\uparrow} & e_{R_{1}}^{r \uparrow} & e_{R_{1}}^{g \uparrow} & e_{R_{1}}^{b \uparrow} & \overline{\nu}_{L_{1}}^{\uparrow} & \overline{u}_{L_{1}}^{\overline{r} \uparrow} & \overline{u}_{L_{1}}^{\overline{b} \uparrow} & \overline{u}_{L_{1}}^{\overline{g} \uparrow}\\
d_{R_{2}}^{\uparrow} & e_{R_{2}}^{r \uparrow} & e_{R_{2}}^{g \uparrow} & e_{R_{2}}^{b \uparrow} & \overline{\nu}_{L_{2}}^{\uparrow} & \overline{u}_{L_{2}}^{\overline{r} \uparrow} & \overline{u}_{L_{2}}^{\overline{b} \uparrow} & \overline{u}_{L_{2}}^{\overline{g} \uparrow}\\
\end{bmatrix}_{Mass}
\end{eqnarray}
Similarly, the second family will be represented by $Q$ and the third family by $R$, both present as spin up and spin down. The three mass families with different spins that transform according to $SU(3)_{grav} \times SU(2)_{R} \times U(1)_{g}$ can be represented as:-
\begin{equation}
    \{ P^{\uparrow} , P^{\downarrow} , Q^{\uparrow} , Q^{\downarrow} , R^{\uparrow} , R^{\downarrow} \}
\end{equation}
\subsubsection{Triality Operator}
 The action of the triality operator on $Cl(8)$ representations \cite{baez2002octonions, lounesto_2001, Stoica_2018} can be seen as:-
\begin{equation}
    Trial : \bigl\{V_{8}, S_{8}^{+} , S_{8}^{-}\bigl\} \longrightarrow \bigl\{S_{8}^{+} , S_{8}^{-}, V_{8}\bigl\}
\end{equation}
\begin{equation}
Trial : \begin{pmatrix}
A^{\uparrow} & 0\\
0 & B^{\downarrow}
\end{pmatrix} = \begin{pmatrix}
C^{\uparrow} & 0\\
0 & A^{\downarrow}
\end{pmatrix}
\end{equation}
where $\{ A ,B , C\} $ represents the usual $SU(2)_{L}$ active generations. 
Now look at the $Cl(9)$ algebra. It gives us the spin up and spin down  for both  flavour as well as mass eigenstates; one that transforms according to $SU(2)_{L}$ and the other transforms according to $SU(2)_{R}$.  If $ \{P, Q , R \}$ are the three generations that transform according to $SU(2)_{R}$, then total space for us is:-
\begin{eqnarray}
\begin{pmatrix}
A^{\uparrow}\oplus A^{\downarrow}\\
B^{\uparrow} \oplus B^{\downarrow}\\
C^{\uparrow} \oplus C^{\downarrow}
\end{pmatrix} \oplus 
\begin{pmatrix}
P^{\uparrow} \oplus P^{\downarrow}\\
Q^{\uparrow} \oplus Q^{\downarrow}\\
R^{\uparrow} \oplus R^{\downarrow}
\end{pmatrix}
\end{eqnarray}
\begin{equation}
    O = 1 \otimes Trial
\end{equation}
Now if we operate the operator $O$ on our total space, we can group various mass and flavour families in a given $Cl(9)$ algebra by permuting the rows. This gives us a theoretical framework to construct the CKM matrix.
\section{CKM Matrix Parameters}
Let us focus our attention on one generation that transforms according to $SU(3)_{grav}\times SU(2)_{R} \times U(1)_{g}$. Here we have eight mass eigenstates or the particles in one generation, considering the two particles that transform according to $SU(3)_{grav}$. Here we develop some isomorphisms to make further progress. As we already had octonionic representations of various particles, and quaternionic representations of particles  from one generation; it was natural to proceed with them. However, those methods did not yield any significant progress, which forced us to adopt the method given below.  
\subsection{Gravi-Charge Operator}
We can develop an isomorphism from the space of representations (space of ideals) of one generation of mass eigenstates to a 
8-dimensional complex vector space. For some definite spin, suppressing the spin, we can write the above argument of isomorphism for all the particles for three generations as given below:-
\begin{equation}
    \{ P , Q , R\} \longrightarrow \mathbb{C}^{8}
    \oplus \mathbb{C}^{8} \oplus \mathbb{C}^{8}
\end{equation}
We can now act on this space of $ \mathbb{C}^{8}
    \oplus \mathbb{C}^{8} \oplus \mathbb{C}^{8}$ with an operator $G$- the \textbf{Gravi - Charge Operator}, to assign the gravi charges to various particles.
\begin{equation}
    G = M_{8}(\mathbb{C})\oplus M_{8}(\mathbb{C})\oplus M_{8}(\mathbb{C})
\end{equation}
\begin{equation}
M_{8}(C) = 
    \begin{bmatrix}
     g_{1}   & 0 & 0 & 0 & 0\\
     0 & g_{2} & 0 & 0 & 0\\
     0 & 0 & g_{3} & 0 & 0\\
     0 & 0 & 0 & g_{4} & 0\\
      0 & 0 & 0 & 0 & --
    \end{bmatrix}
\end{equation}
This matrix $M_{8}(C)$ will be used three times for three mass families. So the gravi-charge operator only has diagonal entries. It acts on linear column vectors that are $SU(2)_{R}$ mass eigenstates and assign them a gravi charge.
\subsection{Mass and Gravi-Charge}
Now before moving further, we make some assumptions:-
\begin{itemize}
    \item Mass is a derived quantity. gravi-charge is more fundamental.
    \item The mass operator will be constructed from the gravi-charge operator, and the gravi-charge eigenvectors are weighed accordingly by the value of the square root of the mass of respective particles to make them massive eigenvectors.
\end{itemize}
\subsection{Left Handed Quarks}
We now look only at a part of the operator $G$ and its action on down, charm and strange quarks, and similarly, the action on up, charm and top quarks.
The operator $G$ can be reduced to a small matrix representation as given below:-
\begin{equation}
G=
    \begin{pmatrix}
       g & 0 & 0\\
       0 & g & 0\\
       0 & 0 & g
    \end{pmatrix}
\end{equation}
It acts on $SU(2)_{R}$ active mass eigenstates and gives the gravi-charges.
\begin{equation}
    \{ u , c , t \} \in \mathbb{C}^{3}
\end{equation}
\begin{equation}
    \{ d , s , b \} \in \mathbb{C}^{3}
\end{equation}
Observe that $\mathbb{C}^{3}$ vector space is needed for both families of quarks.
\begin{figure}[ht!]
    \centering
    \includegraphics[width=8cm]{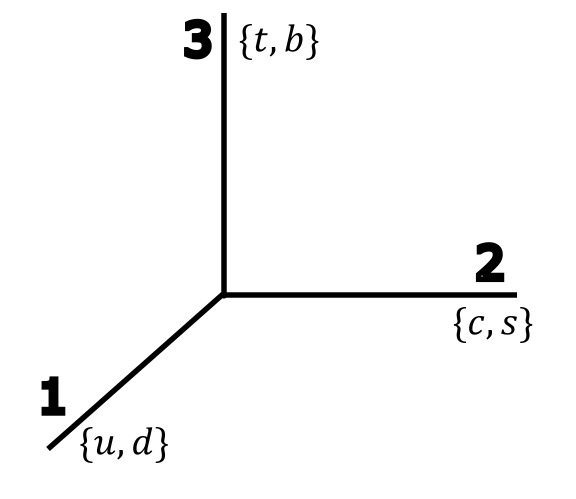}
    \caption{ Basis vectors of reduced vector space $\mathbb{C}^3$ act as $SU(2)_{R}$ active quarks. The space is used for $SU(2)_{L}$ active quarks. }
    \label{fig:example_figure}
\end{figure}\\
We will use this later, when one axis will represent one quark from the up isospin family and one from the down isospin family. This is done to observe the transformation between quark states.
The right handed up quarks (eigenstates of the gravi charge operators) are given by:-
\begin{equation}
    u_{g,R} =  \begin{pmatrix}
       1\\
       0\\
       0
    \end{pmatrix} \:\:\:\:\:\:\:\:\:\ c_{g,R} =  \begin{pmatrix}
       0\\
       1\\
       0
    \end{pmatrix} \:\:\:\:\:\:\:\:\:\ t_{g,R} = \begin{pmatrix}
       0\\
       0\\
       1
    \end{pmatrix} 
\end{equation}
We can define massive quark vectors as :-
\begin{equation}
    u_{m,R} =  \begin{pmatrix}
       \sqrt{m_{u}}\\
       0\\
       0
    \end{pmatrix} \:\:\:\:\ c_{m,R} =  \begin{pmatrix}
       0\\
       \sqrt{m_{c}}\\
       0
    \end{pmatrix} \:\:\:\:\ t_{m,R} = \begin{pmatrix}
       0\\
       0\\
       \sqrt{m_{t}}
    \end{pmatrix} 
\end{equation}
Now in nature, we see a left handed quark, an $SU(2)_{L}$ active left handed quark vector is present. We propose that it is a linear combination of massive quark vectors. So a normalised left handed vector can be represented by :
\begin{equation}
    e_{q}^{\prime} = \frac{1}{\sqrt{m_{u} + \alpha^2m_{c} + \beta^2m_{t}}}\begin{pmatrix}
        \sqrt{m_{u}}\\
        \alpha\sqrt{m_{c}}\\
        \beta\sqrt{m_{t}}
    \end{pmatrix}
\end{equation}
By varying $\alpha$ and $\beta$, we can change the contribution of various massive vectors to the given $SU(2)_{L}$ active left handed quark vector. The same can be done for the down-quark family. However, it should be kept in mind that only the integer linear combination of massive quark vectors can be done. 
\subsection{CKM Matrix}
Now observe these two left handed vectors:-
\begin{equation}
    e_{1}^{\prime} = \frac{1}{\sqrt{m_{u} + \alpha^2m_{c} + \beta^2m_{t}}}\begin{pmatrix}
        \sqrt{m_{u}}\\
        \alpha\sqrt{m_{c}}\\
        \beta\sqrt{m_{t}}
    \end{pmatrix}
\end{equation}
\begin{equation}
    e_{2}^{\prime} = \frac{1}{\sqrt{m_{d} + a^2m_{s} + b^2m_{b}}}\begin{pmatrix}
        \sqrt{m_{d}}\\
        a\sqrt{m_{s}}\\
        b\sqrt{m_{b}}
    \end{pmatrix}
\end{equation}
We try a set of values $\alpha = \beta = a = b = 1$. With this choice, for $e_{1}$ the probability of it being in a top quark gravi eigenstate is $99.33 \% $. Similarly for $e_{2}$ the probability of it being in bottom quark gravi eigenstate will then be equal to $97.7 \%$. So let us identify $e_{1}$ and $e_{2}$; as left handed top quark ($e_{t}$) and a left handed bottom quark ($e_{b}$), respectively.
Now let us see the decay of flavour eigenstate of bottom quark to a flavour eigenstate of top quark : $e_{b}^{\prime} \longrightarrow e_{t}^{\prime}$.
\begin{equation}
    e_{t}^{\prime} = \frac{1}{\sqrt{m_{u} + m_{c} + m_{t}}}\begin{pmatrix}
        \sqrt{m_{u}}\\
        \sqrt{m_{c}}\\
        \sqrt{m_{t}}
    \end{pmatrix}\:\:\:\:\:\
    e_{b}^{\prime} = \frac{1}{\sqrt{m_{d} + m_{s} + m_{b}}}\begin{pmatrix}
        \sqrt{m_{d}}\\
        \sqrt{m_{s}}\\
        \sqrt{m_{b}}
    \end{pmatrix}
\end{equation}
These vectors can be rotated into each other by the application of normal rotation matrices.
\begin{figure}[ht!]
    \centering
    \includegraphics[width=5cm]{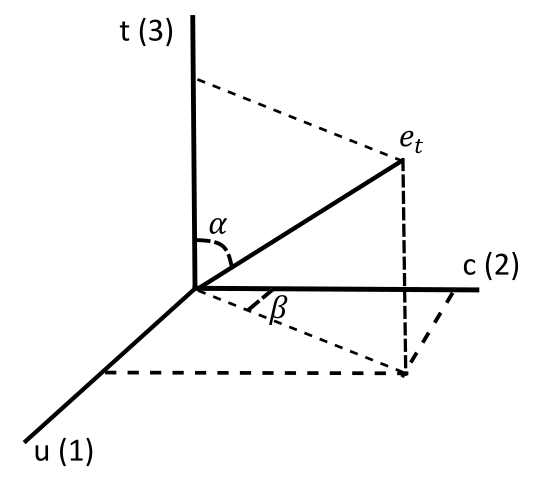}
    \includegraphics[width=5cm]{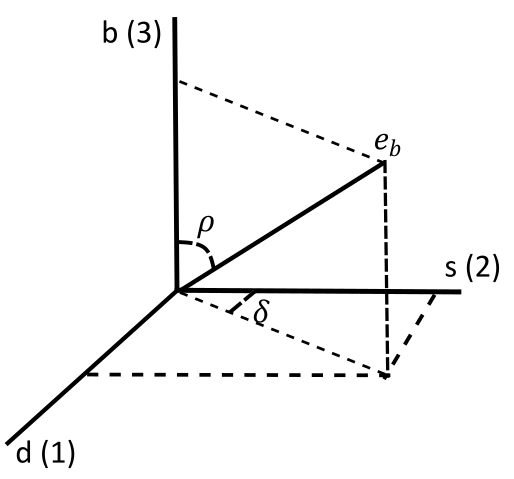}
    \caption{$SU(2)_{L}$ active particles and their projections }
\end{figure}
Here $u$ represents the matrices acting on vectors in the space of the up-isospin particles, and similarly $d$ represents the matrices acting on the space of down-isospin particles.
\begin{equation}
    e_{t}^{\prime} = R_{12}^u(-\beta)R_{23}^u(-\alpha)R_{23}^d(\rho)R_{12}^d(\delta)e_{b}^{\prime} = V e_{b}^{\prime}
\end{equation}\\
\begin{equation}
     R_{12}^d(\delta) = \begin{pmatrix}
     cos(\delta) & -sin(\delta) & 0\\
     sin(\delta) & cos(\delta) & 0\\
     0 & 0 & 1
    \end{pmatrix}
\:\:\:\:\:\:\:\:\:\:\:\:\:\:\:\:\:\:\:\:\:\:\:\:\:\:\:\:\:\:\:\:\:\:\:\:\:\:\:\:\:\:\:\:\:\:\:\:\:\:\:\:\:\:\:\:\:   cos(\delta) = \frac{\sqrt{m_{s}}}{\sqrt{m_{s} + m_{d}}}
\end{equation}\\
\begin{equation}
    R_{23}^d(\rho) = \begin{pmatrix}
        1 & 0 & 0\\
        0 & cos(\rho) &  -sin(\rho)\\
        0 & sin(\rho) & cos(\rho)
    \end{pmatrix}\:\:\:\:\:\:\:\:\:\:\:\:\:\:\:\:\:\:\:\:\:\:\:\:\:\:\:\:\:\:\:\:\:\:\:\:\:\:\:\:\:\:\:\:\:
    cos(\rho) = \frac{\sqrt{m_{b}}}{\sqrt{m_{b}+m_{s}+m_{d}}}
\end{equation}
\begin{equation}
     R_{12}^u(-\beta) = (R_{12}^u(\beta))^T =\begin{pmatrix}
     cos(\beta) & sin(\beta) & 0\\
     -sin(\beta) & cos(\beta) & 0\\
     0 & 0 & 1
    \end{pmatrix}\:\:\:\:\:\:\:\:\:\:\:\:\:\:\:\:\:\:\:\
    cos(\beta) = \frac{\sqrt{m_{c}}}{\sqrt{m_{u} + m_{c}}}
\end{equation}
\begin{equation}
    R_{23}^u(-\alpha) = (R_{23}^u(\alpha))^T = \begin{pmatrix}
        1 & 0 & 0\\
        0 & cos(\alpha) &  sin(\alpha)\\
        0 & -sin(\alpha) & cos(\alpha)
    \end{pmatrix}\:\:\:\:\:\:\:\:\:\:
    cos(\alpha) = \frac{\sqrt{m_{t}}}{\sqrt{m_{u}+m_{c}+m_{t}}}
\end{equation}
Now we use the numerical values of the square root masses of various quarks obtained from the eigenvalues of $3\times 3$ octonionic Hermitian matrices, as shown in Fig. 2. By that substitution, we obtain:-
\begin{equation}
    V_{ij} =  \begin{pmatrix}
     0.9813 & -0.1924 & -0.0030\\
     0.1917 & 0.9789 & -0.0707\\
     0.0165 & 0.0688 & 0.9975
    \end{pmatrix}
\end{equation}
\begin{equation}
    |V_{ij}| =  \begin{pmatrix}
     0.9813 & 0.1924 & 0.0030\\
     0.1917 & 0.9789 & 0.0707\\
     0.0165 & 0.0688 & 0.9975
    \end{pmatrix}
\end{equation}
The code used to obtain the above CKM matrix using square root mass as projections; is presented in the Appendix B.
Every element of $V_{ij}$ represents a projection of quark $j$ on quark $i$. Its square represents the probability of transitioning from quark $j$ to quark $i$ in standard particle physics.
\subsubsection{Standard CKM Matrix}
In standard QFT textbooks \cite{doi:https://doi.org/10.1002/9783527618460.ch10}, it is given that the CKM matrix is just a unitary transformation from mass eigenstates to states that are weak iso-spin doublets. The weak isospin doublets are $SU(2)_{L}$ active.
The weak interaction doublets are given below :
\begin{equation}
    \begin{pmatrix}
        u\\
        d^{\prime}
    \end{pmatrix} \:\:\:\:\:\:\:\:\:\:\:\:\
    \begin{pmatrix}
        c\\
        s^{\prime}
    \end{pmatrix}\:\:\:\:\:\:\:\:\:\:\:\:\
    \begin{pmatrix}
        t\\
        b^{\prime}
    \end{pmatrix}
\end{equation}
The CKM matrix can then be written as:-
\begin{equation}
    \begin{pmatrix}
        d^{\prime}\\
        s^{\prime}\\
        b^{\prime}
    \end{pmatrix} = 
    \begin{pmatrix}
        V_{ud} & V_{us} & V_{ub}\\
        V_{cd} & V_{cs} & V_{cb}\\
        V_{td} & V_{ts} & V_{tb}
    \end{pmatrix}
    \begin{pmatrix}
        d \\
        s\\
        b
    \end{pmatrix}
\end{equation}
The $\{ d , s , b\}$ represents the mass eigenstates. Each entry in the CKM matrix written as $V_{ij}$ represents the transition of $j$ quark to $i$ quark by weak interactions. The CKM matrix is parameterised using three Euler angles $ \{ \theta_{12} , \theta_{13} , \theta_{23} \}$ and a phase factor $\delta_{13}$ \cite{PhysRevLett.53.1802} as given below:-
\begin{equation}
    \begin{pmatrix}
        c_{12}c_{13} & s_{12}c_{13} & s_{13}e^{-i\delta_{13}}\\
        -s_{12}c_{23}-c_{12}s_{23}s_{13}e^{i\delta_{13}} & c_{12}c_{23}-s_{12}s_{23}s_{13}e^{i\delta_{13}} & s_{23}c_{13}\\
        s_{12}s_{23}-c_{12}c_{23}s_{13}e^{i\delta_{13}} & -c_{12}s_{23}-s_{12}c_{23}s_{13}e^{i\delta_{13}} & c_{23}c_{13}
    \end{pmatrix}
\end{equation}
The experimental determination of the entries of the CKM matrix gives the values \cite{10.1093/ptep/ptaa104}:-
\begin{equation}
    \begin{pmatrix}
        |V_{ud}| & |V_{us}| & |V_{ub}|\\
        |V_{cd}| & |V_{cs}| & |V_{cb}|\\
        |V_{td}| & |V_{ts}| & |V_{tb}|
    \end{pmatrix} =
    \begin{pmatrix}
        0.97370 \pm 0.00014 & 0.2245 \pm 0.0008 & 0.00382 \pm 0.00024 \\
        0.221 \pm 0.004 &  0.987 \pm 0.011 & 0.0410 \pm 0.0014\\
        0.0080 \pm 0.0003 & 0.0388 \pm 0.0011 & 1.013 \pm 0.030
    \end{pmatrix}
\end{equation}
This yields the following experimentally determined values of the angles and the complex phase \cite{AMSLER20081} :-
\begin{equation}
    \theta_{12} = 13.04\degree \pm 0.05\degree
\end{equation}
\begin{equation}
    \theta_{13} = 0.201\degree \pm 0.011\degree
\end{equation}
\begin{equation}
    \theta_{23} = 2.38\degree \pm 0.06\degree
\end{equation}
\begin{equation}
    \delta_{13} = 68.8\degree \pm 4.5\degree
\end{equation}
\subsubsection{Theoretical Determination of CKM matrix angles}
With the values of the CKM matrix obtained from the theoretical considerations, we calculated the following values of the CKM Euler angles:-
\begin{equation}
    \theta_{12} = 11.093\degree
\end{equation}
\begin{equation}
    \theta_{13} = 0.172\degree
\end{equation}
\begin{equation}
    \theta_{23} = 4.054\degree
\end{equation}
We have no information about phase in our analysis so far. Further assumptions and research is required in this direction.  The values obtained are in reasonable agreement with the measured values.  Basically, the off-diagonal matrix elements are  different from the experimentally determined values and hence are the reason for these values of the angles. A correction to mass matrices and hence to the masses of particles itself, is required to obtain better values. This is because we have used mass ratios derived in the asymptotically free limit, whereas mixing angles are likely impacted by the running of masses.
\subsubsection{CKM parameters using mass as projections}
Instead of using the square root mass as the projections, we tried using mass. With this new definition our $SU(2)_{L}$ active particles will be given by :-\\
\begin{equation}
    e_{b}^{\prime} \longrightarrow e_{t}^{\prime}
\end{equation}
\begin{equation}
    e_{t}^{\prime} = \frac{1}{\sqrt{m_{u}^2 + m_{c}^2 + m_{t}^2}}\begin{pmatrix}
        m_{u}\\
        m_{c}\\
        m_{t}
    \end{pmatrix}
\end{equation}
\begin{equation}
    e_{b}^{\prime} = \frac{1}{\sqrt{m_{d}^2 + m_{s}^2 + m_{b}^2}}\begin{pmatrix}
        m_{d}\\
        m_{s}\\
        m_{b}
    \end{pmatrix}
\end{equation}\\
\begin{equation}
    e_{t}^{\prime} = R^{u}_{12}(-\beta_{1})R_{23}^{u}(-\beta_{2})R_{23}^{d}(\alpha_{2})R_{12}^{d}(\alpha_{1})e_{b}^{\prime} = Ve_{b}^{\prime}
\end{equation}
We use the same machinery, and rotate the vectors into each other by application of rotation matrices. It gives us following matrix required for the transformation :-
\begin{equation}
V_{ij}=    \begin{pmatrix}
        0.9984 & -0.0559 & 0.2228 \times 10^{-5}\\
        0.0559 & 0.9982 & 0.1236 \times 10^{-2}\\
        -0.7134 \times 10^{-5} & -0.1234 \times 10^{-2} & 0.9998
    \end{pmatrix}
\end{equation}
The code to obtain the above given CKM matrix is presented in the Appendix C.
With the above values of the various CKM matrix elements we obtain the following values of the CKM parameters :
\begin{equation}
    \theta_{12} = 3.205\degree
\end{equation}
\begin{equation}
    \theta_{13} = 0.00013\degree
\end{equation}
\begin{equation}
    \theta_{23} = 0.071\degree
\end{equation}
The above values are very different from  experimentally obtained values. This thus provides us with additional  justification for using the square root mass values over the mass values; while constructing the massive and the $SU(2)_{L}$ active left handed vectors.

\subsubsection{Connection between mass eigenstates and weak-isospin doublets}
Observe that the physically massive vectors used in the above calculations are a linear combination of gravi-charge eigenstates of the right handed quarks. Also, observe that as we have developed an isomorphism between the vector space of ideals to this new vector space $\mathbb{C}^{8}
    \oplus \mathbb{C}^{8} \oplus \mathbb{C}^{8}$, for $SU(2)_{R}$ active mass eigenstates, we can do a similar mapping for the space of the $SU(2)_{L}$ active flavour eigenstates. So for the three left handed quarks of same colour of $SU_{c}(3)$, we will need the following space to describe them $\mathbb{C}
    \oplus \mathbb{C} \oplus \mathbb{C}$, just as for mass eigenstates. This time however, instead of the Gravi-charge operator, another diagonal operator corresponding to the electric charge will act on this space. Let us use the same $\mathbb{C}^{3}$ for both left and right active states (suppressing the colour for both $SU(3)_{c}$ and $SU(3)_{grav}$). Then we can interpret the CKM matrix as a transformation that rotates the normalised mass eigenstates of the gravi-charge vectors to the normalised left handed flavour eigenstates. This connection can be done because of the triality. Triality allows for the mixing of various families in the spinor representations of  the $Cl(8)$ algebra.
\begin{equation}
    \begin{pmatrix}
        d_{L}\\
        s_{L}\\
        b_{L}
    \end{pmatrix} = \begin{pmatrix}
        V_{ud} & V_{us} & V_{ub}\\
        V_{cd} & V_{cs} & V_{cb}\\
        V_{td} & V_{ts} & V_{tb}
    \end{pmatrix}\begin{pmatrix}
        d_{g,R}\\
        s_{g,R}\\
        b_{g,R}
    \end{pmatrix}
\end{equation}
We have to use the normalised mass eigenstates and hence the gravi-charge eigenvectors.
\section{Summary and Discussion}
As is evident from the analysis in the previous sections, the complex Clifford algebra $Cl(9)$ is one of great significance. It is the algebra of unification of the standard model with gravitation, via a left-right symmetric extension of the standard model. We also note that $Cl(9)$ has dimension $512$, and its irrep is $16\times 16$ matrices with complex number entries. If we assume the diagonal entries of these matrices to be real, their dimensionality is reduced to $512-16 = 496$, which is precisely the dimension of the $E_8 \times E_8$ symmetry group $(248+248)$ proposed by us earlier for unification \cite{kaushik2022e8}. Hence there is consistency between $E_8\times E_8$ symmetry and the algebra $Cl(9)$ vis a vis unification. Prior to left-right symmetry breaking which breaks unification in this theory, the coupling constant is simply unity, and the role of the emergent $U(1)$ charge is played by this coupling constant divided by $3$. Thus the fundamental entities prior to symmetry breaking are lepto-quark states which all have an associated charge $1/3$: these are neither bosonic nor fermionic in nature, and the charge value $1/3$ is evident when one finds the eigenmatrices corresponding to the Jordan eigenvalues in the exceptional Jordan eigenvalue problem. For these eigenmatrices  see the Appendix in \cite{bhatt2022majorana}. The neutrino family, the up quark family, the down quark family and the electron family, all are expressed as different superpositions of three basis states which all have an associated charge $1/3$. This means that the left-chiral families are electric charge eigenstates expressed as superposition of pre-unification basis states, and right-chiral families are square-root mass eigenstates expressed as superposition of pre-unification basis states. This fact permits electric charge eigenstates to be expressed as superpositions of square-root mass eigenstates which in turn allows mass ratios to be determined theoretically \cite{Singhfsc}.

We recall from above that the unification algebra $Cl(9)$ is written as a direct sum of two copies of $Cl(8)$. On the other hand $Cl(9)$ can also be written as $Cl(9)= Cl(7)\otimes Cl(2) = [Cl(6)\otimes Cl(2)] \oplus [Cl(6) \otimes Cl(2)]$.
This last expression has profound implications for our understanding of space-time structure in quantum field theory. Recall that each of the two $Cl(6)$ represents one generation of standard model chiral quarks and leptons; the first $Cl(6)$ for left-chiral particles and the second $Cl(6)$ for right-chiral particles. In so far as the $Cl(2)$ are concerned, the second $Cl(2)$ (associated with right chiral fermions)  is used to generate the Lorentz algebra $SL(2,C)$ of 4D space-time (via complex quaternions with one quaternionic imaginary kept fixed), which includes the Lorentz boosts and the three-dimensional $SU(2)_R$ rotations. Gauging of this $SU(2)_R$ symmetry can be used to achieve Einstein's general relativity on a 4D space-time manifold \cite{Woit}. As for the first $Cl(2)$, the one associated with left-chiral fermions, the $SU(2)_L$ rotations  describe weak isospin. However, undoubtedly, this $Cl(2)$ has its own set of Lorentz boosts, which along with the weak isospin rotations generate a second 4D spacetime algebra $SL(2,C)$ distinct from the first, familiar 4D spacetime. In spite of its counterintuitive nature this second spacetime is also an element of physical reality, and there is definitive evidence for it in our earlier work \cite{Singhgrav, Singhexceptional, Sherry}. In this second space-time, distances are at most of the order of the range of the weak force, and only microscopic quantum systems access this second space-time. Classical systems do not access it - their penetration depth into this space-time is much less than one Planck length. Our universe thus has two 4D space-times, which have resulted from the symmetry breaking of a 6D space-time, consistent with the equivalence $SL(2,\mathbb H)\simeq SO(1,5)$. See also \cite{Chester, Tr1, Tr2, Tr3}. The second space-time also obeys the laws of special relativity, and has a causal light-cone structure. A quantum system travels from a space-time point $A$ to another space-time point $B$ through both space-times, but gets to $B$ much faster through the second space-time, on a time scale of the order $L/c\sim 10^{-26}$ s where $L\sim 10^{-16}$ cm is the range of the weak force. This is true even if $B$ is located billions of light years away from $A$, and this offers a convincing resolution of the EPR paradox as to how quantum influences manage to arise nonlocally. These influences are local through the second space-time. In spirit our resolution could be compared to the ER=EPR proposal, but unlike the latter, our resolution has a sound mathematical basis. Moreover our resolution was not invented with the express purpose of understanding quantum nonlocality but is an indirect implication of the algebraic unification of standard model with gravitation. The weak force is seen as the geometry of this second space-time.\\

\noindent{\it How is the Coleman-Mandula theorem evaded by our proposed unification of spacetime and internal symmetries?} The Coleman-Mandula theorem \cite{CM}  is a no-go theorem which states that the space-time symmetry (Lorentz invariance) and internal symmetry of the S-matrix can only be combined in a trivial way, i.e. as a direct product. However, this does not prevent the $E_8\times E_8$ unification of gravitation and the standard model, on which the analysis of the present paper is based. This is because, as pointed out for instance in Section 7 of the work on gravi-weak unification \cite{NP}  the theorem applies only to the spontaneously broken phase, in which the Minkowski metric is present. The unified phase does not have a metric, and hence not the Minkowski metric either, and hence the Coleman-Mandula theorem does not apply to the unified symmetry.\\

\noindent{\it Interpreting the theoretically derived mass ratios:} In the first paragraph of this section we explain how the eigenvalues and eigenmatrices of the exceptional Jordan algebra determine quantization of mass and charge. Furthermore, the expression of charge eigenstates as superposition of mass eigenstates permits derivation of the mass ratios, because mass measurements are eventually carried out using electric charge eigenstates. This explains the strange observed mass ratios of elementary particles. Nonetheless, it is known that masses run with the energy scale, and one can legitimately ask how the derived mass ratios are to be interpreted? The answer is straightforward: the ratio is of those mass values which are obtained in the no-interaction (asymptotically free) limit. Thus the ratio of muon to electron mass has been derived in the low-energy limit, whereas the ratio of say the down quark to electron mass is obtained by comparing the down quark mass at the relatively high energy at which quark asymptotic freedom is achieved, to the electron mass at the low energy free limit. These two compared masses (down-quark and electron) are {\it not} at the same energy. Moreover, all these mass ratios will run with energy - that running is not part of the present derivation, and is left for future work.\\

\noindent{\it Evidence for a second 4D spacetime:} The Clifford algebra associated with the complex quaternions (when none of the quaternionic imaginary directions is kept fixed) is $Cl(3)$, and is a direct sum of two $Cl(2)$ algebras, which together correspond to complex split biquaternions \cite{vaibhav2021leftright}. The spacetime associated with $Cl(3)$ is 6D spacetime $SO(1,5)$ because of the homomorphism $SL(2,H)\sim SO(1,5)$ whereas each of the $Cl(2)$ is individually associated with a 4D spacetime each, because $Cl(2)$ generates the Lorentz algebra $SL(2,C)$. See also the related work of Kritov \cite {Kritov}. The construction of two copies of such a spacetime is made explicit in Eqn. (13) and the subsequent discussion in \cite{Singhgrav} and also in \cite{Singhexceptional}. The presence of a second spacetime is also fully evident in \cite{Sherry} where we have discussed in detail the bosonic content of the spontaneously broken $E_8 \times E_8$ symmetry. \\

\noindent {\it Implications for fundamental physics in the early universe / high-energy regime}: In our algebraic approach to unification, Clifford algebras and the standard model have been studied, with dynamics given by the theory of trace dynamics. The main advantage of this approach is that the spinor representations of the fundamental fermions can be constructed easily here as the left ideals of the algebra. This formalism makes unique predictions for fundamental physics, including new particle content which should be looked for in experiments. The predicted particles include three right handed sterile neutrinos (the only new fermions predicted beyond the standard model), a second (electrically charged) Higgs, eight gravi-gluons associated with the newly predicted $SU(3)_{grav}$ symmetry, and the dark photon associated with the new $U(1)_{grav}$ symmetry which possibly underlies Milgrom's MOND as an alternative to dark matter. We predict that the Higgs bosons are composites of those very fermions to which they are said to assign mass. Prior to electroweak symmetry breaking the universe obeys the unified $E_8 \times E_8$ symmetry which combines the standard model forces with gravitation. In this phase there is no distinction between spacetime and matter, and the fundamental degrees of freedom are the so-called atoms of space-time-matter.

\bigskip
\bigskip
\noindent\textbf{Acknowledgements}\\
We would like to thank Abineet Parichha, for his valuable feedback and suggestions.
\section{Appendix A}
The $3\times 3$ Hermitian Octonionic Matrices, known as the exceptional Jordan algebra satisfy the characteristic equation given as \cite{dray1998octonionic, harvey1990spinors} :-
 \begin{equation}
     A^{3} - (trA)A^{2} + \sigma(A) A - (det A)I = 0
 \end{equation}
 For the definition of each part look at an example shown here.
 \begin{equation}
     A = \begin{pmatrix}
         p & a & \overline{b}\\
         \overline{a} & m & c\\
         b & \overline{c} & n
     \end{pmatrix}
 \end{equation}
 \begin{equation}
     p,m,n \in \mathbb{R} \:\:\:\:\:\:\:\:\:\:\  a,b,c \in \mathbb{O}
 \end{equation}
 \begin{equation}
     tr A = p + m + n
 \end{equation}
 \begin{equation}
     \sigma(A) = pm + pn + mn - |a|^{2} - |b|^{2} - |c|^{2}
 \end{equation}
 \begin{equation}
     det A = pmn + b(ac) + \overline{b(ac)} - n|a|^{2} - m|b|^{2} - p|c|^{2}
 \end{equation}
 The real eigenvalues of the $3\times 3$ Hermitian Octonionic matrix satisfy a modified characteristic equation given by:-
 \begin{equation}
     det(\lambda I - A) = \lambda^{3} - (tr A)\lambda^{2} + \sigma(A)\lambda - det(A) = r
\end{equation}
\begin{equation}
    r^{2} + 4\Phi(a,b,c)r - |[a,b,c]|^{2} = 0
\end{equation}
\begin{equation}
    \Phi(a,b,c) = \frac{1}{2}Re([a,\overline{b}]c)
\end{equation}
\begin{equation}
    [a,b,c] = (ab)c - a(bc)
\end{equation}
The $[a,b,c]$ is the associator it is a  measure of the associativity of the algebra involved. Now for our case the mass matrix has only quaternionic entries. In that case $r=0$, and we have the usual characteristic equation that gives us real roots. These real roots are then used to calculate the mass ratios \cite{bhatt2022majorana}.
\newpage
\section{Appendix B}
Here in this code we use  mass eigenstates weighted by square-root of mass. The method is explained in the section $8.4$. The identifications used in the code are written below :\\
\begin{equation}
    e_{t}^{\prime} = R_{12}^u(-\beta)R_{23}^u(-\alpha)R_{23}^d(\rho)R_{12}^d(\delta)e_{b}^{\prime} = V e_{b}^{\prime}
\end{equation}\\
\begin{equation}
    A12T \longrightarrow R_{12}^{u}(-\beta)
\end{equation}
\begin{equation}
    A23T \longrightarrow R_{23}^{u}(-\alpha)
\end{equation}
\begin{equation}
    B23 \longrightarrow R_{23}^{d}(\rho)
\end{equation}
\begin{equation}
    B12 \longrightarrow R_{12}^{d}(\delta)
\end{equation}
\
\begin{figure}[ht!]
    \centering
    \includegraphics[width=17cm]{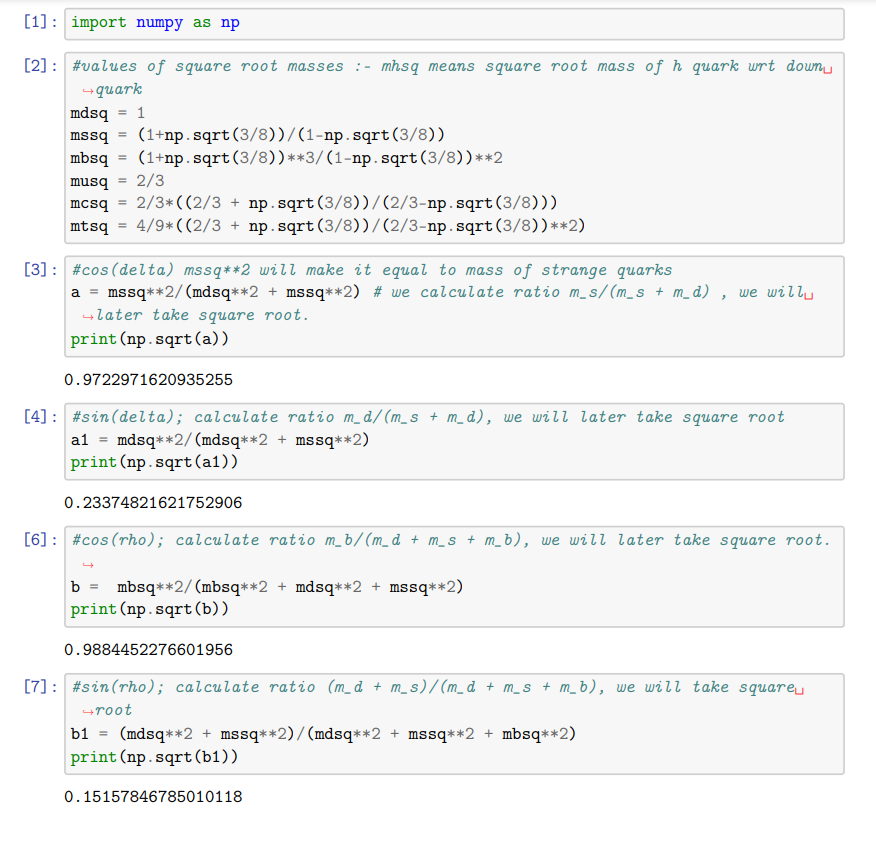}
\end{figure}
\begin{figure}[ht!]
    \centering
    \includegraphics[width=17cm]{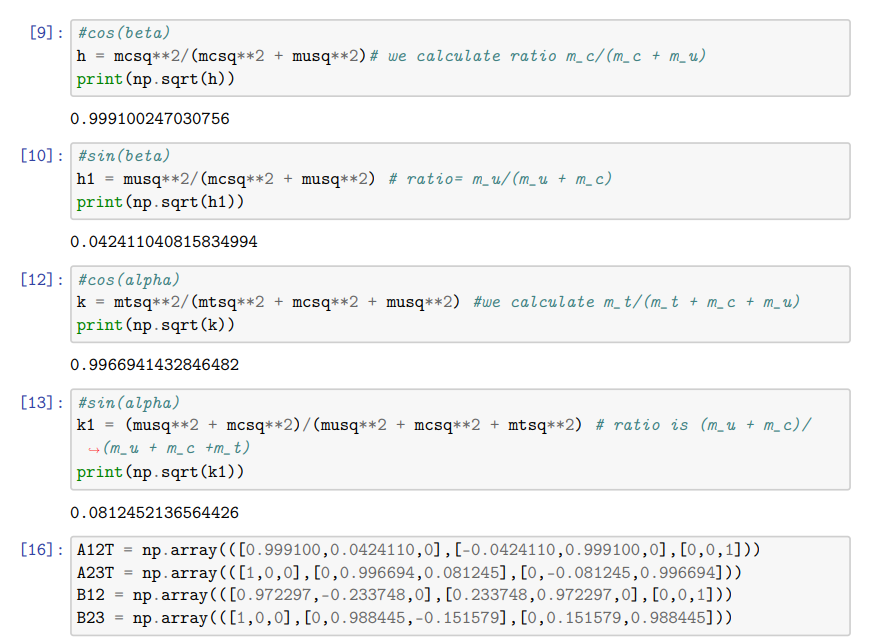}
    \includegraphics[width=17cm]{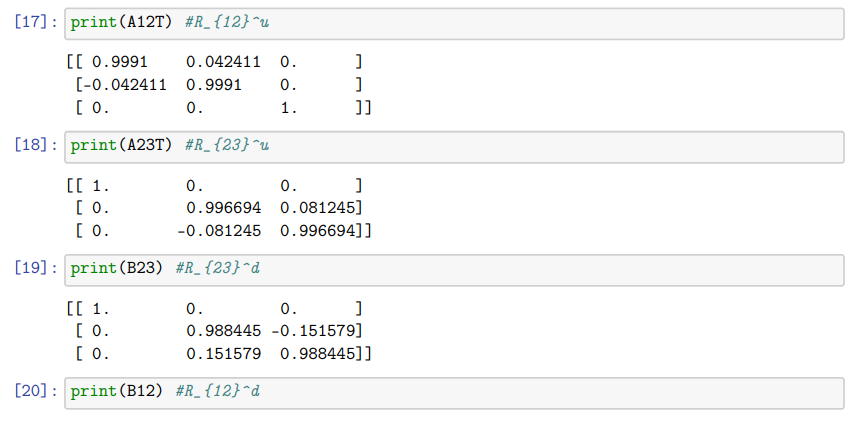}
\end{figure}
\clearpage
\newpage
\begin{figure}[ht!]
    \centering
    \includegraphics[width=17cm]{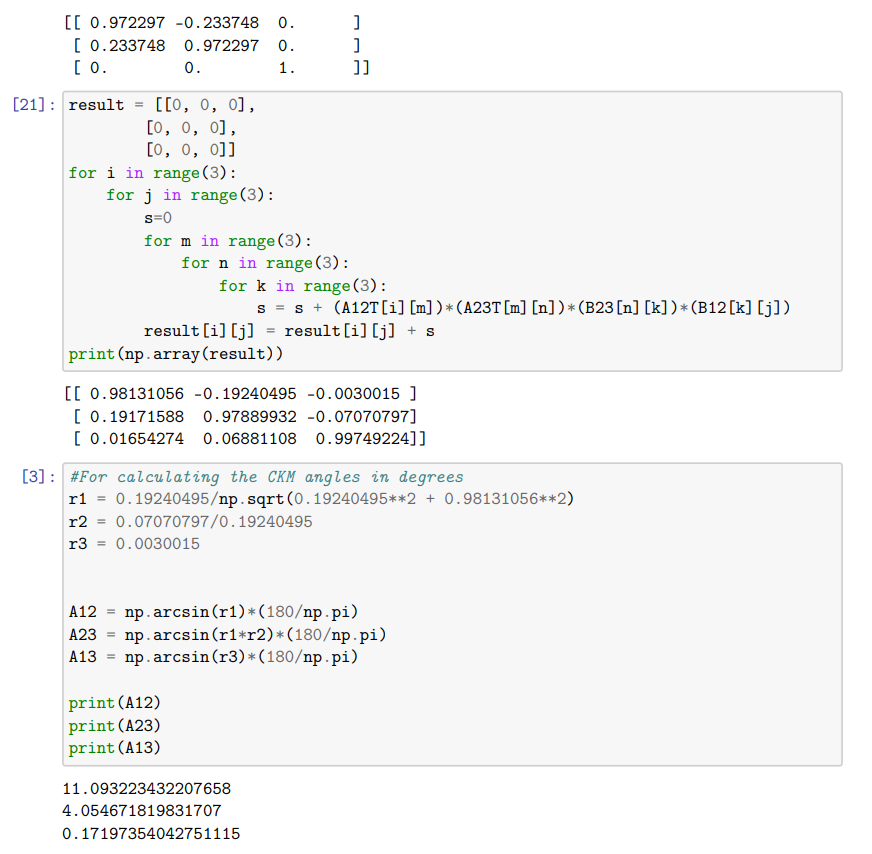}
\end{figure}
These values are reported in the earlier section.
\newpage
\section{Appendix C}
Here is a code for computing CKM matrix parameters and mixing angles with mass eigenstates weighted by mass (instead of square root of mass). The definitions of rotation matrices and the mass vectors correspondingly get changed.\\
\begin{equation}
    e_{t}^{\prime} = R^{u}_{12}(-\beta_{1})R_{23}^{u}(-\beta_{2})R_{23}^{d}(\alpha_{2})R_{12}^{d}(\alpha_{1})e_{b}^{\prime} = Ve_{b}^{\prime}
\end{equation}\\
\begin{equation}
    P12T \longrightarrow R^{u}_{12}(-\beta_{1})
\end{equation}
\begin{equation}
    P23T \longrightarrow R_{23}^{u}(-\beta_{2})
\end{equation}
\begin{equation}
    Q23 \longrightarrow R_{23}^{d}(\alpha_{2})
\end{equation}
\begin{equation}
    Q12 \longrightarrow R_{12}^{d}(\alpha_{1})
\end{equation}
\begin{figure}[ht!]
    \centering
    \includegraphics[width=17cm]{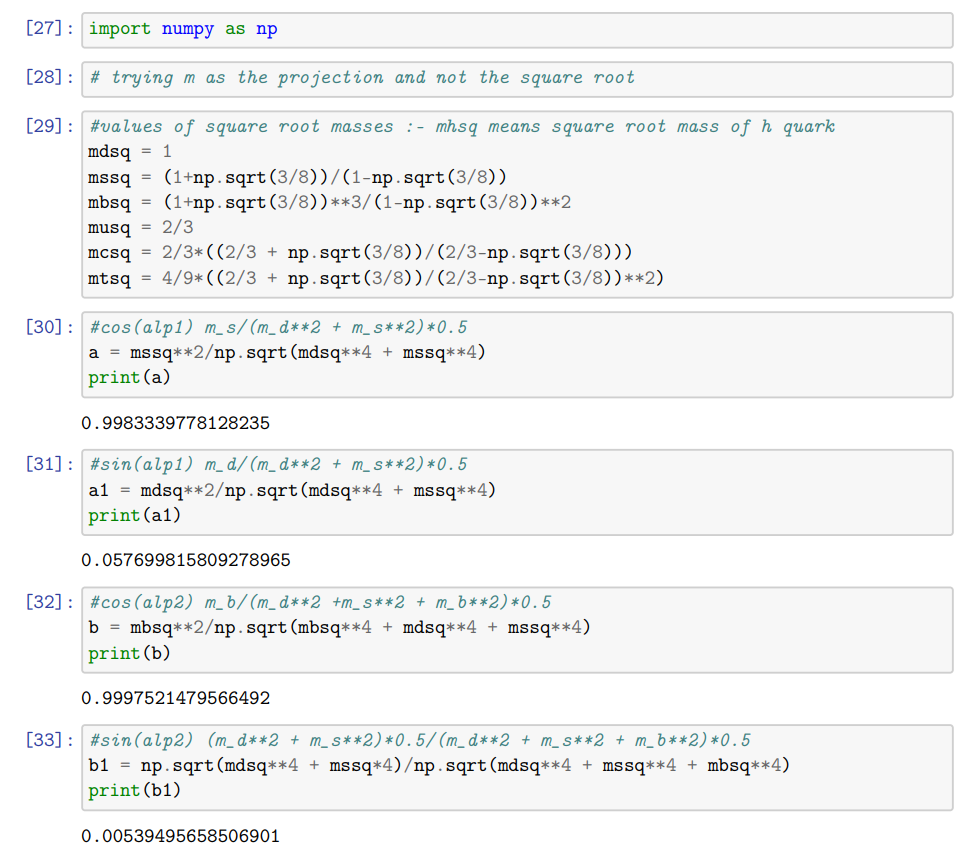}
    \includegraphics[width=17cm]{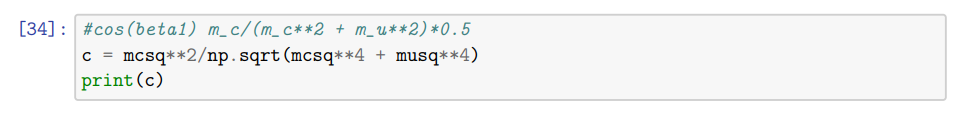}
\end{figure}\\
\clearpage
\newpage
\begin{figure}[ht!]
    \centering
    \includegraphics[width=17cm]{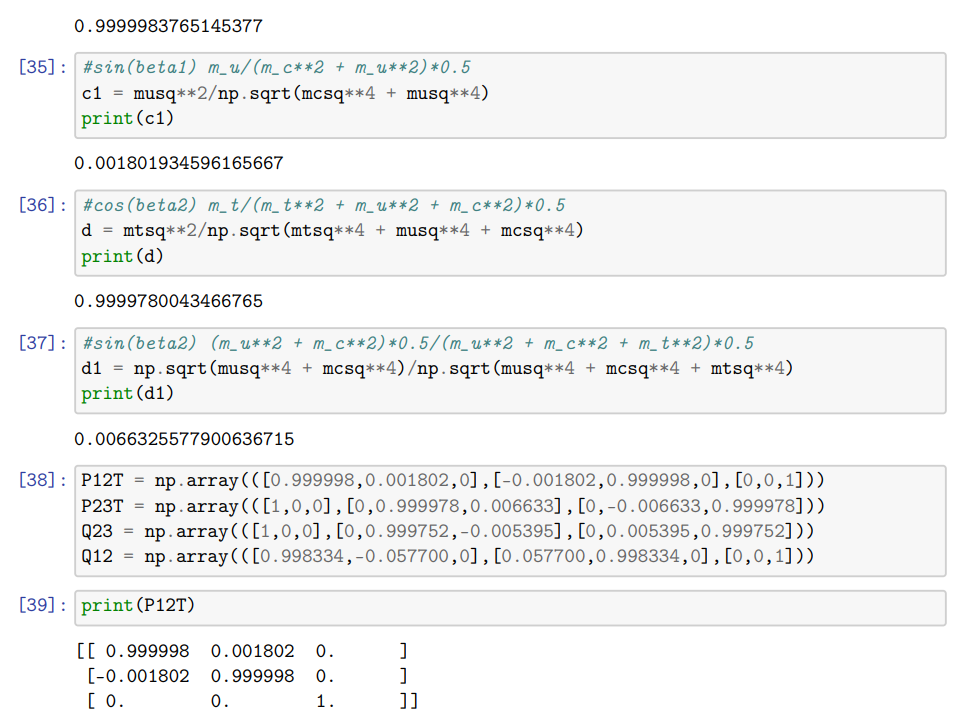}
    \includegraphics[width=17cm]{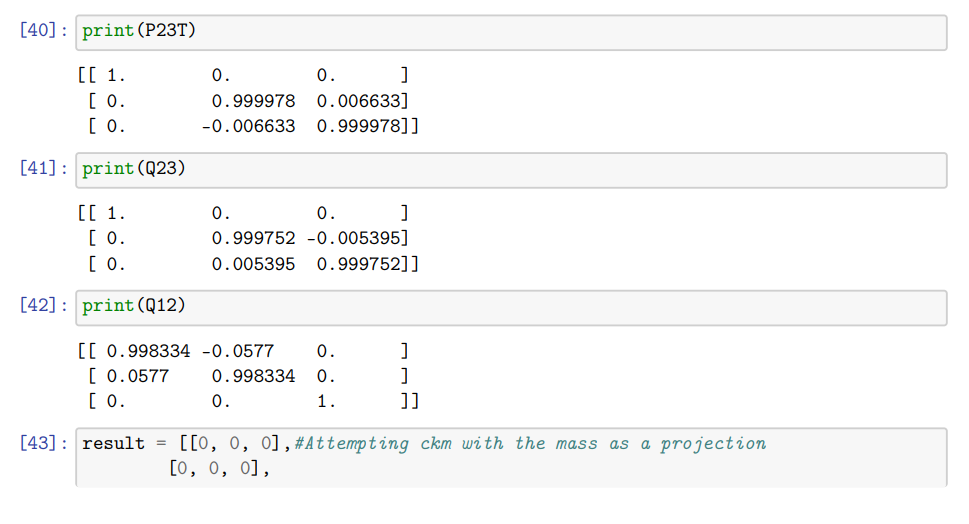}
\end{figure}
\clearpage
\newpage
\begin{figure}[ht!]
    \centering
    \includegraphics[width=17cm]{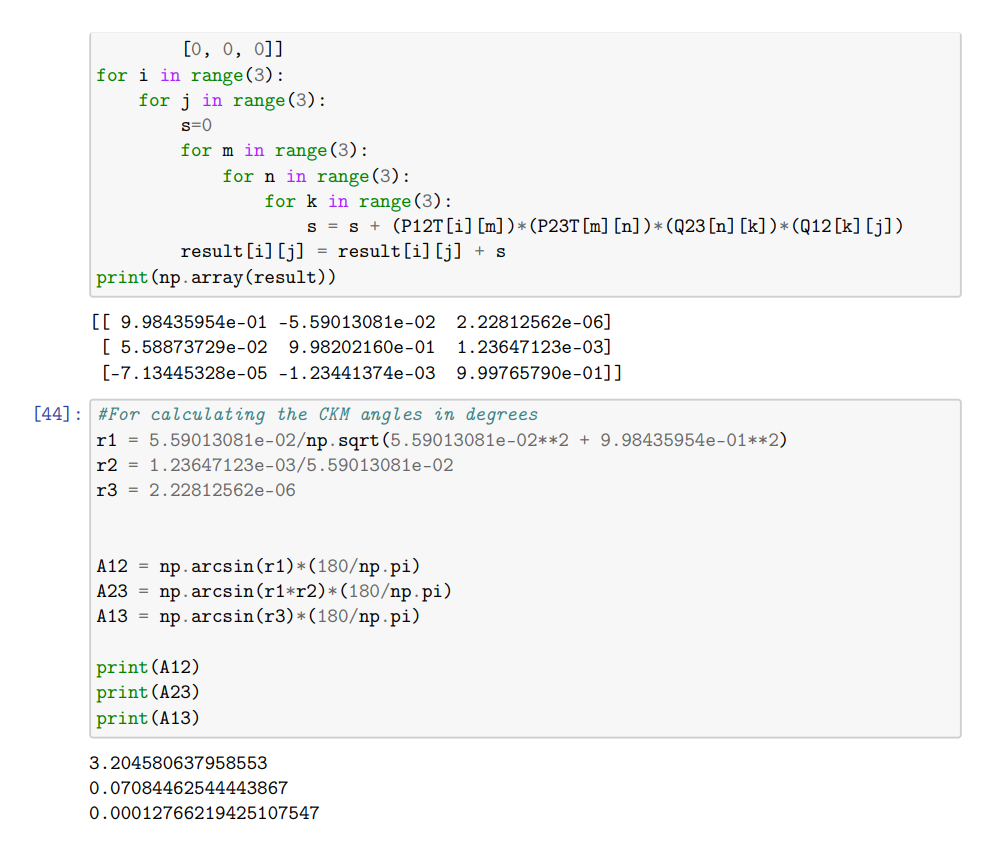}
\end{figure}

Here it can be seen that the values obtained for the CKM parameters are very different from the experimentally seen values. It justifies our choice of using the square root mass as more fundamental quantity over the mass of the fermions.


\end{document}